\documentclass[twocolumn]{aastex63}
\usepackage{courier}
\usepackage{booktabs}
\usepackage{threeparttable}
\usepackage{tabularx}
\usepackage{color}

\shortauthors{Z. Zhuang et al.}

\defcitealias{Leethochawalit2019}{L19}
\defcitealias{Kirby2013}{K13}

\begin{document}

\title{NGC 147 Corroborates the Break in the Stellar Mass-Stellar Metallicity Relation for Galaxies}

\author[0000-0002-1945-2299]{Zhuyun Zhuang}
\affiliation{Department of Astronomy, California Institute of Technology, 1200 E. California Blvd., MC 249-17, Pasadena, CA 91125, USA}

\author[0000-0001-6196-5162]{Evan N.\ Kirby}
\affiliation{Department of Astronomy, California Institute of Technology, 1200 E. California Blvd., MC 249-17, Pasadena, CA 91125, USA}

\author[0000-0003-4570-3159]{Nicha Leethochawalit}
\affiliation{School of Physics, The University of Melbourne, Parkville, VIC 3010, Australia}
\affiliation{ARC Centre of Excellence for All Sky Astrophysics in 3 Dimensions (ASTRO 3D), Australia}
\affiliation{National Astronomical Research Institute of Thailand (NARIT), MaeRim, Chiang Mai, 50180, Thailand}

\author[0000-0002-4739-046X]{Mithi A.\ C.\ de los Reyes}
\affiliation{Department of Astronomy, California Institute of Technology, 1200 E. California Blvd., MC 249-17, Pasadena, CA 91125, USA}

\correspondingauthor{Zhuyun Zhuang}
\email{zzhuang@astro.caltech.edu}
\received{2021 May 28}
\revised{2021 July 2}
\accepted{2021 July 8}

\begin{abstract}
    The stellar mass-stellar metallicity relation (MZR) is an essential approach to probing the chemical evolution of galaxies. It reflects the balance between galactic feedback and gravitational potential as a function of stellar mass. However, the current MZR of local dwarf satellite galaxies ($M_{\ast } \lesssim  10^8 M_{\odot}$, measured from resolved stellar spectroscopy) may not be reconcilable with that of more massive galaxies ($M_{\ast } \gtrsim  10^{9.5} M_{\odot}$, measured from integrated-light spectroscopy). Such a discrepancy may result from a systematic difference between the two methods, or it may indicate a break in the MZR around $10^{9} M_{\odot}$. To address this question, we measured the stellar metallicity of NGC 147 from integrated light using the Palomar Cosmic Web Imager. We compared the stellar metallicity estimates  from integrated light with the  measurements from resolved stellar spectroscopy and found them to be consistent within 0.1~dex. On the other hand, the high-mass MZR overpredicts the metallicity by 0.6~dex at the mass of NGC 147.  Therefore, our results tentatively suggest that the discrepancy between the low-mass MZR and high-mass MZR should not be attributed to a systematic difference in techniques.  Instead, real physical processes cause the transition in the MZR\@. In addition, we discovered a positive age gradient in the innermost region and a negative metallicity gradient from the resolved stars at larger radii, suggesting a possible outside-in formation of NGC 147. 

\end{abstract}


\section{Introduction} \label{sec:intro}
In the past few decades, numerous works have revealed the tight correlation between the stellar masses (luminosities) and the metallicities of galaxies, i.e., the stellar mass-metallicity relation (MZR) \cite[e.g.][]{McClure1968,Lequeux1979,Tremonti2004,Lee2006,Kirby2013,Leethochawalit2018}. The observed MZR indicates that more massive galaxies are more metal-rich than less massive ones.  The trend can be interpreted as a result of metal retention, wherein high-mass galaxies with deep gravitational potential wells can resist galactic winds and retain more metals \citep{Dekel1986}. Recent works have argued the metallicity may also be regulated by star formation efficiency \cite[e.g.,][]{Calura2009,Magrini2012} as well as by the interplay between inflow, outflow, and enrichment \citep{Finlator2008}. Generally, galactic metallicities are measured in two forms: gas-phase and stellar-phase. While gas-phase metallicity reflects the metals in the interstellar medium (ISM) at the time of observation, stellar metallicity indicates the amount of metals incorporated into the stars at the time they formed, which is less susceptible to  instantaneous fluctuations and  represents the metal abundance averaged over the star formation history (SFH)\@. Therefore, the stellar MZR is a more stable indicator of the chemical evolution of galaxies than the gas-phase MZR\@.

The MZR\footnote{In this paper, MZR refers to the stellar-phase MZR, if not otherwise specified.} at $z\sim 0$ can be measured from two independent techniques. Local dwarf galaxies can be resolved into stars, so their stellar metallicities are derived from spectroscopy of individual stars \cite[e.g.,][]{Kirby2013}. On the other hand, more massive galaxies are too far for us to resolve the stars.  Therefore, metallicities are obtained from the integrated-light spectra of whole galaxies \cite[e.g.,][]{Gallazzi2005,Zahid2013,Leethochawalit2018,Leethochawalit2019}. Although the low-mass and high-mass MZRs both indicate more massive galaxies are more metal-rich, they follow different forms of expression. \citet[][hereafter \citetalias{Kirby2013}]{Kirby2013} used \citeauthor{Geha2010}'s (\citeyear{Geha2010}) Keck/DEIMOS spectra to derive the low-mass MZR from resolved stellar spectroscopy in Local Group dwarf galaxies ($M_* \lesssim 10^8 M_{\odot}$)\footnote{\citetalias{Kirby2013}\ measured the stellar metallicities of both quiescent and star-forming dwarf galaxies, but they found no difference in the MZR between these two groups.} as [Fe/H] $\propto M_{\ast}^{0.30\pm 0.02}$, while \citet[][hereafter \citetalias{Leethochawalit2019}]{Leethochawalit2019} measured the high-mass MZR from integrated-light spectra of local massive quiescent galaxies ($M_* \gtrsim  10^{9.5} M_{\odot}$)  as [Fe/H] $\propto M_{\ast}^{0.11\pm 0.02}$.  \citet{Choi2014} similarly observed a flat slope in the high-mass MZR, where the metallicities for massive quiescent galaxies ($M_* \ge 10^{10.6}M_{\odot}$) in the local universe were measured from the stacked spectra in bins of stellar mass. \citet{Sybilska2017} estimated the stellar metallicities for 258 nearby massive quiescent galaxies  ($M_* \gtrsim 10^{9.8}M_{\odot}$) in the ATLAS$^{\rm 3D}$ survey \citep{Cappellari2011} using spectrophotometric indices. They also recovered a high-mass MZR similar to that of \citetalias{Leethochawalit2019}. All these studies indicate a shallower slope and higher normalization in the high-mass MZR than in the low-mass MZR, such that an extrapolation of the high-mass MZR to low mass would predict much higher metallicities than are observed in low-mass galaxies.

One possible explanation for the discrepancy is that there is a physical break in the MZR around $10^{8}-10^{10} M_{\odot}$.  This behavior would mimic the gas-phase MZR, which follows a single power law at low to intermediate masses ($10^{6} M_{\odot} \lesssim M_* \lesssim 10^{9.5} M_{\odot}$), and flattens at the high-mass end \citep[e.g.,][]{Blanc2019}. The possible transition may result from different feedback mechanisms of the two mass ranges. Theoretically, it has been suggested that energy-driven winds dominate in low-mass galaxies \citep{Murray2005}, while momentum-driven winds dominate in high-mass galaxies \citep{Hopkins2012}. \citet{Finlator2008} also suggested that the observed gas-phase MZR at $z\sim 2$ for massive galaxies indicates the momentum-driven winds are dominant in the outflows of high-mass galaxies. Such a transition in the slope of the stellar MZR was seen by \citet{Sybilska2017}, who found that the slope around $10^{9.5} M_{\odot}$ steepens significantly as compared to that at the lowest and highest masses. However, the stellar MZR derived by \citet{Sybilska2017} does not go below $10^{9} M_{\odot}$, so it cannot fill the mass gap in the stellar MZR and capture the behavior of the MZR across the full range of galaxy masses. 
The existing measurements in the mass gap are too scarce to constrain the intermediate-mass stellar MZR\@. 

However, the discrepancy in the MZRs may also result from a systematic difference between the two techniques used to measure stellar metallicities. Previous works that tested the consistency between the stellar population parameters---including the metallicities---among various measurement methods have mainly focused on simple systems like globular clusters \cite[e.g.,][]{GonzaezDelgado2010,Conroy2014,Conroy2018}. So far, only \citet{Ruiz-Lara2018} have compared the metallicity of each SFH age bin obtained from the integrated-light spectrum of Leo A, a star-forming dwarf irregular (dIrr) galaxy with $M_{\ast} \sim 3\times 10^6 M_{\odot}$, with that measured by \citet{Kirby2017} from individual stars.  They found that the integrated-light [Fe/H] at all age bins is higher than the resolved-star [Fe/H]\@. The differences in [Fe/H] range from $\sim 0.25$--1.5 dex. However, the differences here can be explained by other reasons, such as the inaccurate modeling of Balmer nebular emission lines, the incompleteness of metal-poor stars in the current stellar libraries, or a negative metallicity gradient, as suggested by \citet{Kirby2017}.

In this work, we focus on NGC 147, a dwarf elliptical (dE) satellite of M31 with $M_{\ast} \simeq 10^8 M_{\odot}$ \citepalias{Kirby2013}. This galaxy is a good target to investigate possible systematic differences between different techniques of measuring stellar metallicities.  First, the galaxy is near enough (712~kpc; \citealt{Conn2012}) to obtain individual stellar metallicities but also far enough to be observed in integrated light with an integral field unit (IFU)\@.  Second, no H$\,${\sc i} gas has been detected in NGC 147, so the spectrum is free from nebular emission line contamination. Third, NGC 147 is more metal-rich than Leo A, so it will suffer less from the incompleteness of metal-poor stars in the current stellar libraries.

The main goals of this paper are (1) to determine whether there is a systematic offset between integrated-light and resolved-star spectroscopic measurements of NGC~147's stellar metallicity, (2) to use these measurements to bolster or diminish the evidence of a break in the stellar MZR, and (3) to leverage IFU data cubes to study the spatial distribution of the stellar population in the innermost region.  We use both new IFU data cubes and archival resolved-star spectra (Section~\ref{sec:data}).  We use a variety of full spectral fitting algorithms (Section~\ref{sec:fitting}) to estimate the global stellar metallicity and age from integrated light (Section~\ref{sec:stellarpops}). Section~\ref{sec:met} specifically addresses goal (1) in determining the consistency between the two different metallicity techniques. We also present some interesting spatial trends of the stellar population that suggest that NGC~147 formed outside in (Section~\ref{sec:spatial}). We compare our work with the previous literature on the stellar metallicity determinations and discuss the subsequent implications for the MZR (goal 2) in Section~\ref{sec:discussion}.    Finally, we summarize our findings in Section~\ref{sec:summary}.


\section{Observations}\label{sec:data}

\subsection{Palomar Cosmic Web Imager Sample}

\subsubsection{Observations and data reduction}
\begin{figure}[t]
    \centering
    \includegraphics[width=0.45\textwidth]{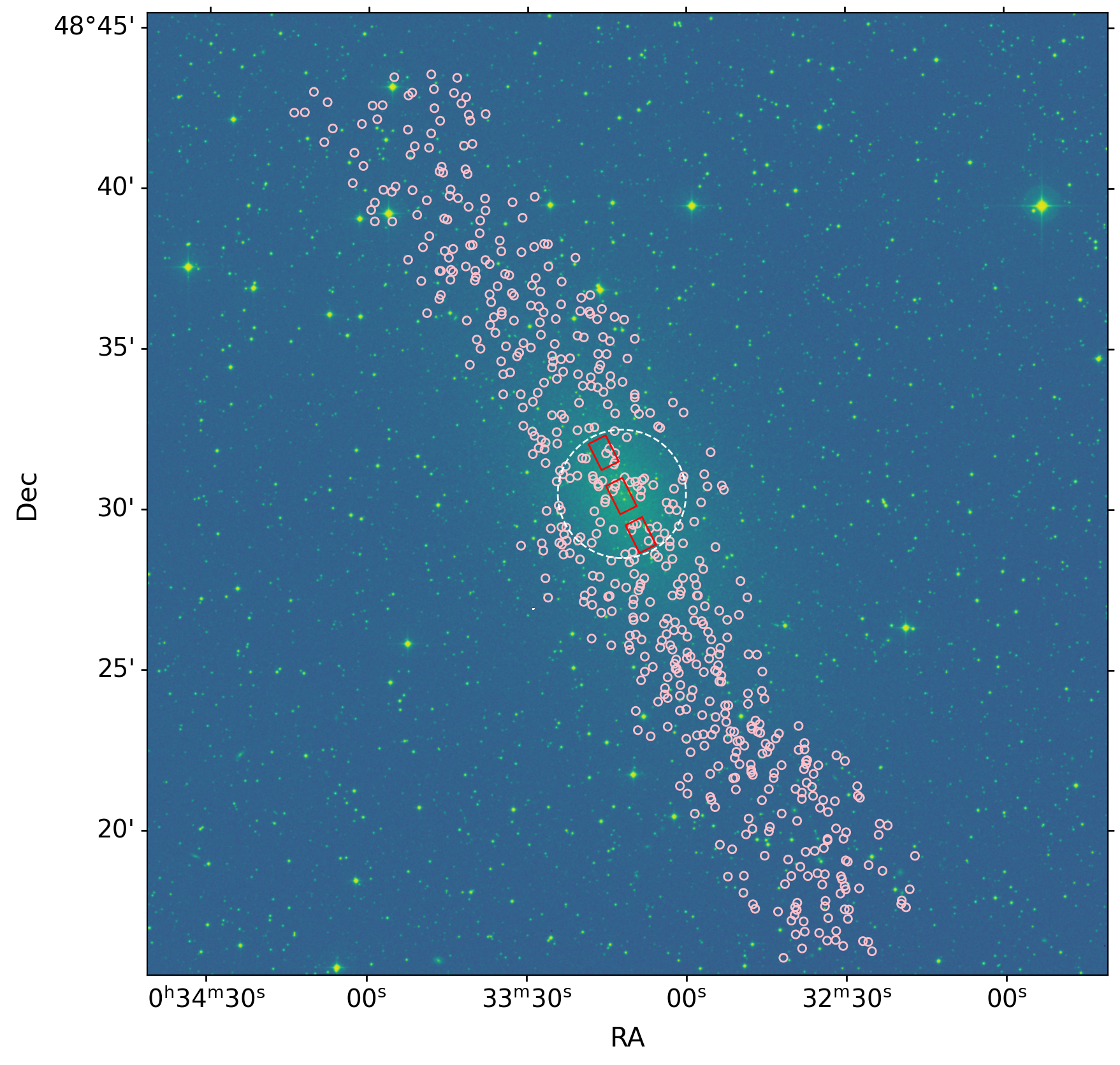}
    \caption{Digitized Sky Survey (DSS) image of NGC 147. The red rectangles show the three PCWI fields observed. The dashed circle indicates the circular effective radius. The pink circles show locations of the RGB star samples used in this study.}
    \label{fig:position}
\end{figure}

Observations for NGC 147 were performed with the Palomar Cosmic Web Imager (PCWI) on 2019 September 23 and 24. PCWI is a medium-resolution integral field spectrograph (IFS) mounted on the Cassegrain focus of the Palomar 200-inch Hale Telescope with a field of view of $60''\times 40''$. It consists of 24 spatial slices with a width of $\sim 2.5''$ and an in-slice pixel scale of $\sim 0.5''$. The seeing was around $2''$ for two nights, so both the spatial axis along each slice and spectral $z$-axis are seeing-limited while the spatial axis across the slices was slit-limited. NGC 147 has an effective radius ($R_{\rm eff}$) of $\sim 6.7'$, which means that a single PCWI pointing covers only a small portion of the galaxy.  We observed the central regions of the galaxy within $2'$ with three pointings (Figure \ref{fig:position}), as described in Table~\ref{tab:observations}. 

\begin{deluxetable*}{ccccccccc}[t]
\tablecaption{Basic Properties of NGC 147}
\label{tab:ngc147_properties}
\tablehead{\colhead{R.A.} & \colhead{Decl.} & \colhead{$M_{\rm V}$} & \colhead{$m_{\rm i,TRGB}$} & \colhead{$d_{\rm MW}$} & \colhead{$R_{\rm eff}$} & \colhead{$\sigma_v$} & \colhead{$\log ({M_{\ast}/M_{\odot}})$} & \colhead{[Fe/H]}\\ \colhead{(J2000)} & \colhead{(J2000)} & \colhead{(mag)} & \colhead{(mag)} & \colhead{(kpc)} & \colhead{(arcmin (kpc))} & \colhead{(km s$^{-1}$)} & \colhead{(dex)} & \colhead{(dex)} }
\startdata
00 33 12.1 & +48 30 31 & -15.33 & $20.82^{+0.08}_{-0.08}$ & $712^{+21}_{-19}$ &  6.70 (1.41) & $16\pm 1$ & $8.00\pm 0.05$ &$-0.83\pm 0.25$
\enddata
\tablecomments{The columns show (1–2) the equatorial coordinates from NED for the galaxy center; (3-6) the absolute total $V$ magnitude, the apparent TRGB $i$-band magnitude, the de-projected distance from the Milky Way (MW) and the effective radius taken from \citet{Crnojevic2014}; (7) the velocity dispersion measured from the resolved stars \citep{Geha2010}; and (8-9) the stellar mass and stellar metallicity obtained from the resolved RGB stars taken from \citetalias{Kirby2013}.}
\end{deluxetable*}

The MEDREZ grating was used to provide the largest spectral range ($\sim 900$\AA) that PCWI offers and a resolution of $R \simeq 1500$. With two configurations centered at $4300$~\AA\  and $5140$~\AA, respectively, the spectra span from $3750$~\AA\ to $5550$~\AA, which is sufficient to measure age, metallicity, and Mg enhancement via full-spectrum fitting \citep{Leethochawalit2018,Leethochawalit2019}. 

Because the galaxy is larger than the field of view of PCWI, additional off-field blank-sky observations were taken adjacent to each science observation with the same exposure time for sky subtraction. The weather was cloudy with the Moon up in the late half of the second night, leading to significant variance in the sky spectra. Therefore, we had to abandon all science frames of the blue side of the north pointing due to inaccurate sky subtraction. Table \ref{tab:observations} lists all the observations used in this study.

\begin{deluxetable*}{ccccccc}[t]
\tablecaption{PCWI Observations of NGC 147}
\label{tab:observations}
\tablehead{\colhead{Pointing} & \colhead{R.A.} & \colhead{Decl.} & \colhead{Surface Brightness$^a$} & \colhead{Central Wavelength} & \colhead{Date} & \colhead{Exposure Time} 
\\ \colhead{} & \colhead{(J2000)} & \colhead{(J2000)} & \colhead{(g-band mag arcsec$^{-2}$)} & \colhead{(\AA)} & \colhead{} & \colhead{(s)}  }
\startdata
    Central & 00 33 12.223 &  +48 30 27.341 & 21.32 &   4300 & 2019 Sep 23 & 6$\times$900 \\
    {} & {} & {} & {} & 5140 &  2019 Sep 23 & 6$\times$900\\
    South & 00 33 08.561 &  +48 29 14.526 & 21.59 & 4300 & 2019 Sep 23 & 4$\times$900\\
    {} & {} & {} & {} & 5140 & 2019 Sep 24 & 5$\times$900\\
    North & 00 33 15.570 & +48 31 48.328 & 21.58 & 5140 & 2019 Sep 24 & 6$\times$900
\enddata
\tablecomments{Some observations lacked quality sufficient for this study.  This table does not include those observations.}
\tablenotetext{a}{Derived from the best-fit surface brightness profile for diffuse light in \citet{Crnojevic2014}.}
\end{deluxetable*}

 The raw data were reduced using the standard PCWI Data Reduction Pipeline (DRP)\footnote{PCWI DRP: \url{https://github.com/scizen9/pderp}}. This pipeline converts raw, 2D science frames into flux-calibrated, 3D cubes with spatial positions and wavelengths. PCWI is subject to gravitational flexure because it is mounted at the Cassegrain focus.  Flexure induces minor changes in the spectrum, such as alteration of the wavelength solution. Flexure correction in the PCWI DRP relies on cross-correlation of a time series of data cubes. However, the spectral offsets were not completely eliminated after the flexure correction. We recalibrated the wavelength solution by cross-correlating the spatially collapsed spectra of each image with the sky spectrum taken nearest to the night calibration set. In this way, the wavelength solution was standardized to a common reference.  We applied only a constant offset and ignored higher-order terms.
 
 The world coordinates of each cube were corrected with \texttt{CWITools} \citep{OSullivan2020b} via comparing the location of known foreground point sources with accurate R.A. and decl. The sky frames were subtracted from the adjacent science frames directly to obtain the sky-subtracted cubes. The sky-subtracted cubes were coadded using \texttt{CWITools}, which calculates the footprint of each input pixel on the coadd frame and distributes flux onto the coadd grid.

The instrumental resolution was measured by fitting the widths of arc lines as a function of wavelength. PCWI uses a ThAr lamp, which gives a dense spectrum of emission lines.  As a result, some arc lines were blended. We restricted the arc line list to isolated arc lines in order to eliminate the bias introduced by blended lines. We identified isolated arc lines by setting thresholds on the first and second derivatives of the arc flux. First, we required that the first derivative must be close to zero and that the second derivative must be negative, in order to find the locations of the line peaks. Second, we required that the peak must be sharp rather than having a flat or broadened feature. This means that the first derivative must abruptly change from positive to negative, i.e., that the second derivative must be smaller than some threshold. We experimented with different combinations of the thresholds and visually inspected the result. We found that setting the first derivative of the spectrum to be smaller than 0.2 and the second derivative to be smaller than $-0.01$ provides relatively clean isolated lines. The FWHMs of the isolated lines were then fitted into a quadratic polynomial to obtain the resolution ($\simeq 3.27$~\AA\ or 90 km s$^{-1}$).

  However, there was still a slight offset between the blue and red spectra, as we only applied first-order correction. Instead of fitting for the higher-order terms, we transformed the observed wavelengths into the rest frame separately for both wavelength settings.
 The spatially collapsed spectra of each cube were extracted from the sky-subtracted cubes. We then cross-correlated them with a template galaxy spectrum that was smoothed to the PCWI resolution and re-sampled to the wavelength grid of our data using \texttt{SpectRes} \citep{Carnall2017}. A template with a simple stellar population (SSP) was generated from the Flexible Stellar Population Synthesis (FSPS) \citep{Conroy2009,Conroy2010}, based on the physical properties of NGC 147 listed in Table \ref{tab:ngc147_properties}. Because NGC 147 has an extended SFH that ended $\sim$ 4 Gyr ago, we adopted a stellar population age of 8 Gyr, which approximately corresponds to the average age of the stellar population \citep{Weisz2014}. We conducted further analyses only on the rest-frame spectra.

\subsubsection{Foreground Star Masking}
As shown in Figure \ref{fig:position}, some bright foreground stars are located in the field of PCWI\@.  We masked them before the spectrum extraction to avoid contamination of the galaxy spectrum. The masking regions were determined from the white-light image. For NGC 147, the tip of the red giant branch has an $i$-band magnitude of $\sim 20.82$ \citep{Crnojevic2014}, so even the brightest star members would be too faint to be detected individually by PCWI\@. Therefore, any resolved stars in the field were considered foreground stars. 
We used the DSS image to identify the foreground stars. For each pointlike source on the DSS image within the PCWI field, we modeled its point spread function (PSF) on the white-light image assuming a 2D elliptical Gaussian profile. The point-source mask was determined as a circle centered at the fitted peak, with a radius of three times the fitted semi-major axis. Nevertheless, some bright residuals, which may be related to scattered light, were still present in the field. We masked all the regions above 1.35 times the median. We also experimented with several thresholds, and it turned out that 1.35 times the median removed most of the bright extended features, which might have contaminated the spectra, while leaving most of the galaxy contribution.

\subsubsection{Spectrum Extraction}\label{subsubsec:spectrum_extraction}
IFS data are known to underestimate the variance due to the covariance between adjacent spaxels. This effect originates from the redistribution of the flux from the same pixel onto a new spatial grid when coadding or smoothing the data cubes. When we coadded the single-exposure cubes, the flux within the same slice with a width of $\sim$ 2.5 arcsec was redistributed again into the final coadded image with a pixel scale of 0.58 arcsec. Therefore, the signal-to-noise ratio (S/N) derived from the PCWI DRP was significantly overestimated. 

To estimate the true noise, we followed the procedures in \citet{OSullivan2020} for PCWI data. First, the variance for each cube was first rescaled by 1.5 to account for the covariance introduced by coadding. For the covariance added by spatially binning the neighboring spaxels, \citet{OSullivan2020} measured the true noise ($\sigma_{\rm adjusted}$) in the binned data as 
\begin{equation}
    \sigma_{\rm adjusted}= [1+0.79\log{(N_b)}]\sigma_{\rm nocov},
\label{eq:sn_calibration}
\end{equation}
where $\sigma_{\rm nocov}$ is the noise derived from error propagation and $N_b$ is the number of spaxels in the binned spectrum. For large $N_b \gg 120$, the signals in different spaxels are uncorrelated, so the scaling becomes $\sigma_{\rm adjusted}\simeq$ 2.6 $\sigma_{\rm nocov}$. 

 We stacked the extracted spectra from all the coadded cubes into a single spectrum, which resembles the long-slit integrated-light spectra of more massive galaxies. To remove sensitivity to possibly imperfect flux calibration, we normalized the spectra by the median of the flux between 4400~\AA\ and 4800~\AA\@. We then scaled them according to the surface brightness profile measured by \citet{Crnojevic2014} (Table~\ref{tab:observations}) and summed them to obtain the global spectrum. In this case, the number of spaxels used for spectrum extraction is much larger than 120. Therefore, we scaled the weighted variance by $2.6^2 = 6.76 $.
 
 In addition, we extracted the spatially resolved spectra using the Voronoi binning algorithm of \citet{Cappellari2003}, which is one of the most commonly used binning methods in IFS data. It can adaptively combine the spectra in adjacent pixels to reach a given minimum S/N, and it allows us to adjust the S/N using using Equation~\ref{eq:sn_calibration} when optimizing the binning strategy.
 We only included the central and south pointings for the spatially resolved studies because the north pointing lacks reliable spectral coverage below 4700~\AA, which is important to determine the stellar population age. 
 For the central and south pointings, the S/N was calculated from the wavelength range between
 4400-4600~\AA\@. We required each bin to have S/N $>$ 20~pix$^{-1}$ (26~\AA$^{-1}$) for the central pointing and S/N  $>$ 15~pix$^{-1}$ (20~\AA$^{-1}$)  for the south pointing. We made this choice because the south pointing has many fewer bins in the field due to its lower surface brightness.
 In the end, we have 17 and 4 bins for the central and south pointings, respectively (Figure~\ref{fig:ssp_maps}).  
 
 We further corrected the variance of the stacked spectrum and the spatially resolved spectra by comparing with the best-fit model in order to obtain accurate uncertainties of the stellar population parameters (see Section \ref{sec:fitting}).

\subsection{Keck/DEIMOS Sample}\label{subsec:Deimos_sample}

We used archival Keck/DEIMOS \citep{Faber2003} observations of four slitmasks along the major axis of NGC~147.  These observations were originally obtained by \citet{Geha2010}.  We downloaded the raw data from the Keck Observatory Archive and reduced them with the \texttt{spec2d} software \citep{Cooper2012,Newman2013}.

\citetalias{Kirby2013} already measured stellar metallicities from coadded spectra of red giants in NGC~147.  However, one of the purposes of this paper is to determine whether individual stellar metallicities are consistent with integrated light.  Spectral coaddition is a form of integrated light, which means that the existing measurements in \citetalias{Kirby2013} are not appropriate for our purpose.  Therefore, we remeasured the stellar metallicities for individual stars.

We measured metallicities by $\chi^2$ fitting to a grid of synthetic spectra, as described by \citet{Kirby2008,Kirby2009,Kirby2010}.  This method uses photometric estimates of effective temperature and surface gravity to guide the fitting.  We used photometry from the Pan-Andromeda Archaeological Survey \citep[PAndAS,][]{Martin2013}, which took place at the Canada--France--Hawaii 3.6~m Telescope (CFHT)\@.  This photometry is different from that used by \citetalias{Kirby2013}, which possibly had errors in its zero-points. Only the stars with uncertainties of [Fe/H] smaller than 0.5~dex were included in the further analysis. This cut gives us 317 stars in the sample. The updated measurements are listed in Table~\ref{tab:deimos}.

\begin{deluxetable*}{cccccccccccc}[t]
\tablecaption{Metallicity Catalog for RGB Stars Observed with Keck/DEIMOS}
\label{tab:deimos}
\tabletypesize{\footnotesize}
\tablehead{ \colhead{ID} & \colhead{R.A.} & \colhead{Decl.} & \colhead{Filter 1} & \colhead{Magnitude 1$^a$} & \colhead{Filter 2} & \colhead{Magnitude 2$^a$} & \colhead{$T_{\rm eff}$} & \colhead{$\log g$} & \colhead{$\xi$} & \colhead{[Fe/H]}\\  \colhead{ } & \colhead{ } & \colhead{ } & \colhead{ } & \colhead{($\mathrm{mag}$)} & \colhead{ } & \colhead{($\mathrm{mag}$)} & \colhead{($\mathrm{K}$)} & \colhead{($\mathrm{cm\,s^{-2}}$)} & \colhead{($\mathrm{km\,s^{-1}}$)} & \colhead{($\mathrm{dex}$)}}
\startdata
4456 & 00 32 23.9 & +48 19 53.8 & $g$ & $23.065\pm0.027$ & $i$ & $20.921\pm0.012$ & $3744\pm 44$ & $0.26\pm 0.1$ & 2.08 & $-1.12\pm0.12$ \\
4608 & 00 32 25.8 & +48 18 57.2 & $g$ & $23.106\pm0.028$ & $i$ & $21.007\pm0.013$ & $3770\pm 42$ & $0.32\pm 0.1$ & 2.07 & $-0.49\pm0.11$ \\
4778 & 00 32 27.6 & +48 18 13.8 & $g$ & $22.715\pm0.020$ & $i$ & $21.117\pm0.014$ & $4126\pm 45$ & $0.62\pm 0.1$ & 2.00 & $-1.10\pm0.12$ \\
4885 & 00 32 29.5 & +48 17 34.2 & $g$ & $23.297\pm0.032$ & $i$ & $20.808\pm0.011$ & $3592\pm 56$ & $0.04\pm 0.1$ & 2.13 & $-3.03\pm0.17$ \\
4894 & 00 32 29.7 & +48 19 03.8 & $g$ & $23.239\pm0.031$ & $i$ & $21.424\pm0.018$ & $3929\pm 41$ & $0.64\pm 0.1$ & 1.99 & $-0.81\pm0.12$ \\
4961 & 00 32 30.5 & +48 19 46.6 & $g$ & $24.125\pm0.066$ & $i$ & $21.426\pm0.018$ & $3528\pm 51$ & $0.23\pm 0.1$ & 2.09 & $-3.21\pm0.23$ \\
5026 & 00 32 31.1 & +48 18 04.2 & $g$ & $22.977\pm0.025$ & $i$ & $21.163\pm0.014$ & $3936\pm 39$ & $0.53\pm 0.1$ & 2.02 & $-0.85\pm0.11$ \\
5129 & 00 32 32.0 & +48 16 37.4 & $g$ & $23.416\pm0.035$ & $i$ & $21.145\pm0.014$ & $3748\pm 52$ & $0.31\pm 0.1$ & 2.07 & $-2.23\pm0.16$ \\
5324 & 00 32 33.6 & +48 16 36.3 & $g$ & $23.099\pm0.027$ & $i$ & $21.238\pm0.015$ & $3923\pm 49$ & $0.54\pm 0.1$ & 2.02 & $-1.08\pm0.13$ \\
5368 & 00 32 33.9 & +48 17 50.8 & $g$ & $22.779\pm0.021$ & $i$ & $21.256\pm0.016$ & $4232\pm 42$ & $0.72\pm 0.1$ & 1.97 & $-1.36\pm0.12$  \\
... & ... & ... & ... & ... & ... & ... & ... & ... & ... & ... 
\enddata
\tablecomments{This table is available in its entirety in the machine-readable format in the online journal. A portion is shown here for guidance regarding its form and content.}
\tablenotetext{a}{PAndAS photometry corrected for extinction.}
\end{deluxetable*}

\section{Methods} \label{sec:fitting}
We used full-spectrum fitting via stellar population synthesis (SPS) to measure the stellar population of NGC 147. Compared to spectrophotometric indices, full-spectrum fitting derives the stellar population properties using the information from the whole spectrum instead of portions with strong absorption lines, so high-precision measurements can be achieved.

 We adopted the fitting method of \citetalias{Leethochawalit2019}, from which the high-mass MZR was derived, and we refer the reader to that paper for a full description.  We chose the same fitting algorithms to minimize the systematic difference caused by different full-spectrum fitting codes and different templates.
 In summary, \citetalias{Leethochawalit2019} used the SSP models from FSPS \citep{Conroy2009,Conroy2010} version 3.0 with the \citet{Kroupa2001} initial mass function (IMF), Padova isochrones \citep{Marigo2007}, and the MILES spectral library \citep{Sanchez-Blazquez2006}. The metallicities and ages of the templates range within $-1.98 \le \log{Z} \le 0.20$ and 0.3 Myr--14 Gyr, respectively. We interpreted [Z/H] in the base models as [Fe/H]. In addition to the stellar metallicity [Fe/H], the enhancements of Mg and N were also included using the theoretical response functions from \citet{Conroy2018}, which depend on age and metallicity. The templates were smoothed to match with the resolution of the observed data. We fit for the velocity dispersion, stellar population age, [Fe/H] and [Mg/Fe] by interpolating the SSP models. Therefore, the obtained ages and metallicities are SSP-equivalent.  \citet{Leethochawalit2018}  tested the SSP assumption by fitting the mock composite stellar population (CSP) model spectra, which have exponentially declining SFHs, with the SSP models, and successfully recovered the true metallicity within 0.1 dex.  This result indicates that SSP-equivalent metallicity agrees well with the light- or mass-weighted metallicity. We emphasize that the goal of this paper is to investigate whether the high-mass MZR measured in SSP-equivalent [Fe/H] has systematic offsets relative to the low-mass MZR. 
 As demonstrated in Appendix~\ref{appendix: comparison}, although the choice of SSP or CSP model will affect the stellar age estimates, the measured metallicity is much less affected. Therefore, adopting SSP models will not affect the major conclusion of this paper. We focus on the recovery of SSP-equivalent parameters in the following analyses.

To avoid a possible mismatch in  flux calibration between the blue and red spectra, we normalized the spectral continuum on both sides using a spline fit. \citetalias{Leethochawalit2019} developed an algorithm to minimize any alteration to the template spectra.  The algorithm applies a synthesized continuum to the observed continuum-normalized spectrum. In the first iteration, we fit the continuum-normalized
observed spectrum with continuum-normalized model spectra, and generated a synthesized continuum by fitting a B-spline to the quotient of the continuum-normalized spectrum and the best-fit SSP model spectrum. Subsequent fits were performed on the corrected observed spectrum with unaltered model spectra. The method is described in full detail in \citetalias{Leethochawalit2019}.

For each spectrum, we performed two separate fits. For the first fit, we only masked out the wavelength regions contaminated by the strong sky line residuals. The Ca$\,${\sc ii} H and K lines were also excluded because of the strong nonlocal thermodynamic equilibrium (NLTE) effects, which the current models poorly reproduce. After the first fit, we masked out the regions that were $\geqslant 3\sigma$ from the best-fit model. We visually inspected these newly masked regions to make sure they deviated from the model due to weak sky features rather than deficiencies in the fitting algorithm. Then, we rescaled the variance derived from Section \ref{subsubsec:spectrum_extraction} by the reduced $\chi^2$ of the first fit to estimate the true noise. We then fit the spectrum with new masking regions and corrected variance again to measure the stellar population properties.

To validate our measurements, we also compared our results with those derived from \texttt{alf} \citep{Conroy2018}, which is another full-spectrum fitting algorithm capable of measuring detailed abundances using the Markov Chain Monte Carlo (MCMC) method of parameter estimation.  We used the ``simple'' mode of \texttt{alf}, which also assumes SSPs. The [Fe/H] and [Mg/Fe] results from our own code are consistent with those obtained from \texttt{alf} within 2$\sigma$.

\begin{figure}[t]
    \centering
    \includegraphics[width=0.45\textwidth]{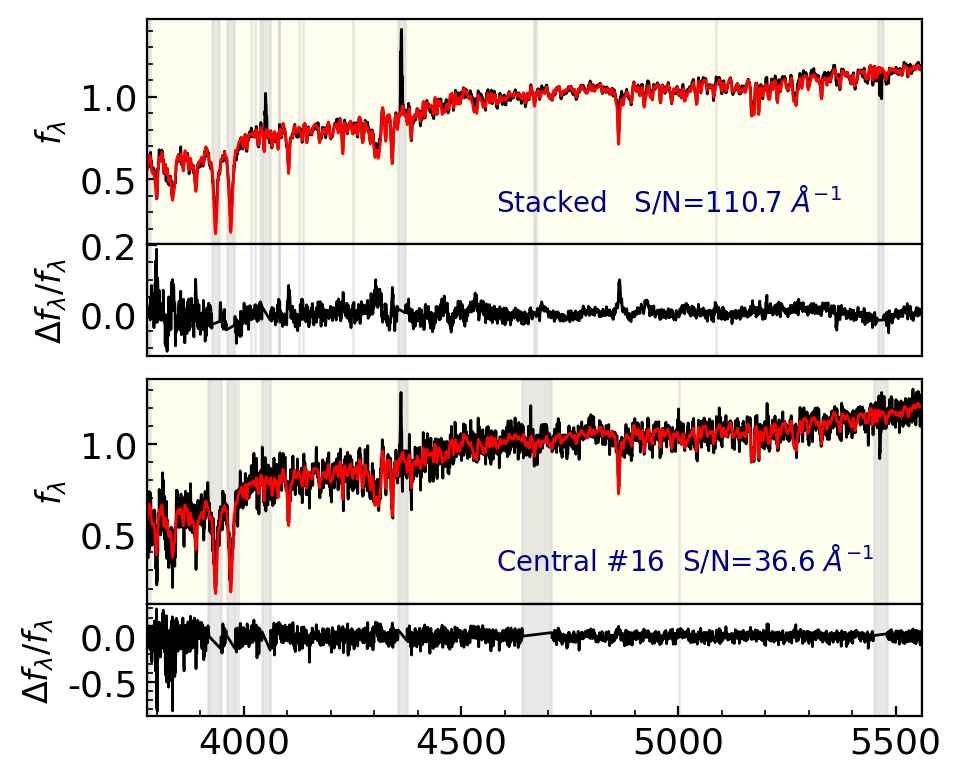}
    \caption{Two examples of the observed spectra (black), the corresponding best-fit models (red), and the model residuals (bottom panel of each spectrum). The gray regions are not included in the spectrum fitting. The top panel shows the stacked spectrum from the entire field, which has the highest S/N, while the bottom panel is an example of a spectrum at the lower end of S/N in one Voronoi cell.  The S/N is estimated from the residual of the fits. }
    \label{fig:fitting_example}
\end{figure}

Examples of the observed spectra and best-fit models at the highest and lowest ends of the S/N are shown in Figure~\ref{fig:fitting_example}. We summarize our results for the global spectrum in Table~\ref{tab:best-fit_stacked} and for the spatially resolved spectra in Table~\ref{tab:best-fit}.

\begin{deluxetable}{cccc}[t]
\tablecaption{Best-fit stellar population parameters for the stacked spectrum of NGC 147 }
\label{tab:best-fit_stacked}

\tablehead{
\colhead{Row} & \colhead{Quantity} & 
\colhead{Units} & \colhead{Best-fit value}
} 
\startdata
(1) & $\sigma_{\rm SSP}$ & km/s & $40.45 \pm  1.85 $ \\
(2) & Age$_{\rm SSP}$ & Gyr & $8.11_{-0.54}^{+0.74}$ \\
(3) & [Fe/H]$_{\rm SSP}$ & dex & $-1.069_{-0.031}^{+0.029}$ \\
(4) & [Mg/Fe]$_{\rm SSP}$ & dex & $0.024_{-0.20}^{+0.20}$ \\
(5) & $\sigma_{\rm alf}$ & km/s & $23.59_{-3.68}^{+3.35}$ \\
(6) & Age$_{\rm alf}$ & Gyr & $12.50_{-0.31}^{+0.32}$ \\
(7) & [Fe/H]$_{\rm alf}$ & dex & $-1.105_{-0.010}^{+0.011}$ \\
(8) & [Mg/Fe]$_{\rm alf}$ & dex & $0.090_{-0.017}^{+0.017}$ \\
\enddata
\tablecomments{(1)-(4) are derived from \citetalias{Leethochawalit2019}'s method; (5)-(8) are obtained from the \texttt{alf} simple mode.}
\end{deluxetable}

\begin{deluxetable*}{ccccccccccc}
\tablecaption{Best-fit stellar population parameters for spatially resolved integrated spectra}
\label{tab:best-fit}
\tablewidth{700pt}
\tabletypesize{\footnotesize}
\tablehead{\colhead{Pointing} & \colhead{R.A.} & \colhead{Decl.} & \colhead{$R_{\rm ellip}$} & \colhead{$\rm Age_{SSP}$$^a$} & \colhead{$\rm[Fe/H]_{SSP}$$^a$} & \colhead{$\rm[Mg/Fe]_{SSP}$$^a$} & \colhead{$\rm Age_{alf}$} & \colhead{$\rm[Fe/H]_{alf}$} & \colhead{$\rm[Mg/Fe]_{alf}$} &
\colhead{S/N$^b$} \\ \colhead{ } & \colhead{(J2000) } & \colhead{(J2000) } & \colhead{(arcmin (kpc))} & \colhead{($\mathrm{Gyr}$)} & \colhead{($\mathrm{dex}$)} & \colhead{($\mathrm{dex}$)} & \colhead{($\mathrm{Gyr}$)} & \colhead{($\mathrm{dex}$)} & \colhead{($\mathrm{dex}$)} &
\colhead{(\AA$^{-1}$)}}
\startdata
Central & 00 33 11.72 & +48 30 33.8 & 0.14 (0.03) & 7.08$^{+0.69}_{-0.44}$ & -1.11$^{+0.03}_{-0.04}$ & 0.13$^{+0.24}_{-0.28}$ & 7.86$^{+0.34}_{-0.43}$ & -0.99$^{+0.04}_{-0.03}$ & 0.14$^{+0.05}_{-0.05}$ & 40.6 \\
{} & 00 33 12.37 & +48 30 40.2 & 0.17 (0.04) & 7.55$^{+0.89}_{-0.55}$ & -1.11$^{+0.04}_{-0.04}$ & 0.44$^{+0.20}_{-0.25}$ & 11.67$^{+0.63}_{-0.52}$ & -1.20$^{+0.03}_{-0.03}$ & 0.35$^{+0.05}_{-0.05}$ & 39.9 \\
{} & 00 33 13.20 & +48 30 35.8 & 0.25 (0.05) & 7.43$^{+0.89}_{-0.54}$ & -1.09$^{+0.04}_{-0.04}$ & -0.04$^{+0.28}_{-0.29}$ & 9.35$^{+0.57}_{-0.70}$ & -1.12$^{+0.04}_{-0.04}$ & 0.09$^{+0.06}_{-0.06}$ & 38.4 \\
{} & 00 33 12.97 & +48 30 27.0 & 0.29 (0.06) & 7.63$^{+1.19}_{-0.68}$ & -1.05$^{+0.04}_{-0.04}$ & -0.37$^{+0.32}_{-0.31}$ & 9.40$^{+0.55}_{-0.39}$ & -0.96$^{+0.03}_{-0.03}$ & -0.11$^{+0.05}_{-0.05}$ & 42.1 \\
{} & 00 33 13.28 & +48 30 48.2 & 0.34 (0.07) & 7.85$^{+1.55}_{-0.88}$ & -1.08$^{+0.04}_{-0.05}$ & -0.18$^{+0.32}_{-0.33}$ & 12.14$^{+0.81}_{-0.81}$ & -1.21$^{+0.03}_{-0.03}$ & -0.03$^{+0.06}_{-0.06}$ & 36.0 \\
{} & 00 33 10.62 & +48 30 20.6 & 0.35 (0.07) & 7.83$^{+1.17}_{-0.71}$ & -1.15$^{+0.04}_{-0.04}$ & -0.36$^{+0.28}_{-0.28}$ & 11.05$^{+0.54}_{-0.41}$ & -1.28$^{+0.03}_{-0.03}$ & -0.10$^{+0.05}_{-0.05}$ & 44.1 \\
{} & 00 33 11.43 & +48 30 42.2 & 0.37 (0.08) & 7.62$^{+0.85}_{-0.54}$ & -1.11$^{+0.04}_{-0.04}$ & 0.09$^{+0.25}_{-0.28}$ & 12.01$^{+0.65}_{-0.64}$ & -1.19$^{+0.03}_{-0.03}$ & 0.10$^{+0.06}_{-0.06}$ & 39.6 \\
{} & 00 33 12.49 & +48 30 15.4 & 0.42 (0.09) & 8.87$^{+1.63}_{-1.12}$ & -1.14$^{+0.05}_{-0.05}$ & -0.33$^{+0.32}_{-0.31}$ & 13.90$^{+0.08}_{-0.35}$ & -1.31$^{+0.03}_{-0.04}$ & 0.01$^{+0.06}_{-0.06}$ & 36.3 \\
{} & 00 33 10.98 & +48 30 08.6 & 0.43 (0.09) & 7.03$^{+0.69}_{-0.42}$ & -1.03$^{+0.03}_{-0.04}$ & -0.42$^{+0.32}_{-0.31}$ & 7.17$^{+0.34}_{-0.28}$ & -1.03$^{+0.03}_{-0.03}$ & -0.02$^{+0.04}_{-0.05}$ & 39.7 \\
{} & 00 33 11.91 & +48 30 10.6 & 0.44 (0.09) & 7.87$^{+1.39}_{-0.88}$ & -1.10$^{+0.04}_{-0.05}$ & -0.09$^{+0.30}_{-0.31}$ & 9.96$^{+0.57}_{-0.56}$ & -1.07$^{+0.03}_{-0.03}$ & -0.05$^{+0.06}_{-0.06}$ & 36.1 \\
{} & 00 33 12.18 & +48 30 53.6 & 0.48 (0.10) & 7.79$^{+1.42}_{-0.86}$ & -1.07$^{+0.04}_{-0.04}$ & -0.05$^{+0.30}_{-0.32}$ & 10.87$^{+0.70}_{-0.60}$ & -1.12$^{+0.04}_{-0.04}$ & 0.17$^{+0.06}_{-0.06}$ & 35.1 \\
{} & 00 33 13.95 & +48 30 30.2 & 0.51 (0.11) & 7.98$^{+1.38}_{-0.92}$ & -1.11$^{+0.04}_{-0.04}$ & 0.14$^{+0.26}_{-0.29}$ & 12.43$^{+0.65}_{-0.70}$ & -1.12$^{+0.03}_{-0.03}$ & 0.05$^{+0.06}_{-0.06}$ & 39.1 \\
{} & 00 33 09.99 & +48 30 11.7 & 0.51 (0.11) & 8.00$^{+1.50}_{-0.92}$ & -1.11$^{+0.05}_{-0.05}$ & -0.19$^{+0.33}_{-0.32}$ & 12.60$^{+0.76}_{-0.89}$ & -1.18$^{+0.03}_{-0.03}$ & -0.06$^{+0.06}_{-0.06}$ & 34.4 \\
{} & 00 33 14.36 & +48 30 42.1 & 0.52 (0.11) & 7.22$^{+0.94}_{-0.58}$ & -1.06$^{+0.04}_{-0.04}$ & -0.01$^{+0.28}_{-0.30}$ & 9.25$^{+0.55}_{-0.47}$ & -1.06$^{+0.04}_{-0.03}$ & 0.07$^{+0.06}_{-0.06}$ & 36.9 \\
{} & 00 33 13.59 & +48 30 21.7 & 0.54 (0.11) & 7.72$^{+1.26}_{-0.80}$ & -1.10$^{+0.04}_{-0.04}$ & -0.09$^{+0.30}_{-0.30}$ & 12.90$^{+0.61}_{-0.69}$ & -1.18$^{+0.03}_{-0.03}$ & 0.02$^{+0.06}_{-0.06}$ & 37.2 \\
{} & 00 33 12.22 & +48 30 01.1 & 0.69 (0.14) & 7.94$^{+1.19}_{-0.73}$ & -1.07$^{+0.04}_{-0.04}$ & -0.04$^{+0.29}_{-0.30}$ & 13.29$^{+0.52}_{-0.78}$ & -1.20$^{+0.03}_{-0.03}$ & 0.13$^{+0.05}_{-0.05}$ & 42.6 \\\hline
South & 00 33 08.32 & +48 29 22.1 & 1.33 (0.28) & 8.92$^{+1.57}_{-1.07}$ & -1.05$^{+0.03}_{-0.04}$ & -0.00$^{+0.25}_{-0.29}$ & 13.93$^{+0.05}_{-0.23}$ & -1.05$^{+0.02}_{-0.02}$ & 0.03$^{+0.05}_{-0.05}$ & 38.7 \\
{} & 00 33 09.93 & +48 29 19.5 & 1.38 (0.29) & 8.36$^{+1.41}_{-0.92}$ & -1.07$^{+0.04}_{-0.04}$ & -0.04$^{+0.29}_{-0.31}$ & 13.56$^{+0.35}_{-0.68}$ & -1.08$^{+0.03}_{-0.03}$ & 0.05$^{+0.05}_{-0.04}$ & 39.9 \\
{} & 00 33 07.47 & +48 29 04.0 & 1.67 (0.35) & 8.73$^{+1.72}_{-1.25}$ & -1.10$^{+0.04}_{-0.04}$ & -0.09$^{+0.29}_{-0.32}$ & 13.46$^{+0.42}_{-0.98}$ & -1.08$^{+0.03}_{-0.03}$ & 0.05$^{+0.05}_{-0.05}$ & 38.2 \\
{} & 00 33 09.16 & +48 28 59.7 & 1.76 (0.37) & 8.56$^{+1.52}_{-0.95}$ & -1.08$^{+0.04}_{-0.04}$ & 0.05$^{+0.26}_{-0.31}$ & 13.81$^{+0.16}_{-0.56}$ & -1.10$^{+0.03}_{-0.03}$ & 0.06$^{+0.04}_{-0.04}$ & 40.0
\enddata
\tablenotetext{a}{Measured from \citetalias{Leethochawalit2019}'s method.}
\tablenotetext{b}{The S/N is calculated from the residuals of the best-fit model.}

\end{deluxetable*}

\section{Global Properties} \label{sec:stellarpops}
In this section, we report the stellar population properties of NGC 147 from the global integrated-light spectrum and compare the best-fit SSP-equivalent results with the updated measurements of the resolved stars.

\subsection{SSP Age}
\label{subsec:ssp_age}

The best-fit SSP age derived from our method is $8.11^{+0.74}_{-0.54}$~Gyr, while \texttt{alf} obtains an age of $12.50^{+0.32}_{-0.31}$~Gyr. We suspect the difference ($\sim0.14$ dex) may be attributed to systematic discrepancies of the SSP models used in the fitting codes. First, \texttt{alf} uses the newer MIST isochrones \citep{Dotter2016,Choi2016} with metallicity coverage larger than that of the Padova isochrones used in our method. Second, the intrinsic spectral resolution of the stellar libraries in \texttt{alf} is lower than the instrumental resolution of PCWI, so the observed spectra needed to be smoothed by another $\sigma=$100~km~s$^{-1}$ before being input to \texttt{alf}. Third, even in the simple mode, \texttt{alf} includes six more elements than our code does. Degeneracy between the extra free parameters may also affect the posterior distributions and the final results.

Ideally, we would compare the age of the stellar population measured from integrated light to the average ages of individual stars.  Unfortunately, it is extremely challenging to measure the ages of individual RGB stars. Generally, the stellar age can only be indirectly estimated by comparing the observed data with stellar models \citep[e.g.,][]{Soderblom2010}. The most commonly used method is to place the observations on a color-magnitude diagram (CMD), and match them with isochrones in a Hertzsprung–Russell diagram. This method works best for main-sequence and subgiant stars and is less effective for old, metal-poor RGB stars. For example, the difference in color between a 9~Gyr and 12~Gyr red giant is on the order of the difference in color between isochrone sets modeling the same star.

We therefore infer the accuracy of the SSP age by comparing it with the age measured from CMD fitting via Hubble Space Telescope (HST) photometry by \citet{Weisz2014}. They obtained a mass-weighted age of 7.53~Gyr, more consistent with the SSP age derived from our algorithm than with that derived from \texttt{alf}. The agreement suggests that even though NGC 147 has an extended SFH with several stellar populations formed at distinct epochs, the SSP assumption may be valid to recover the mass-weighted age in some cases, though the recovered stellar age is highly model-dependent. Therefore, obtaining a more reliable and consistent measurement of stellar
age for galaxies sometimes requires a model of the SFH more
complex than an SSP. As suggested in Appendix~\ref{appendix: comparison}, CSP models may be more appropriate to recover the stellar age of complicated systems, although this method would also require higher data quality.

\subsection{Iron abundance}  \label{sec:met}
\begin{figure}[t]
    \centering
    \includegraphics[width=0.45\textwidth]{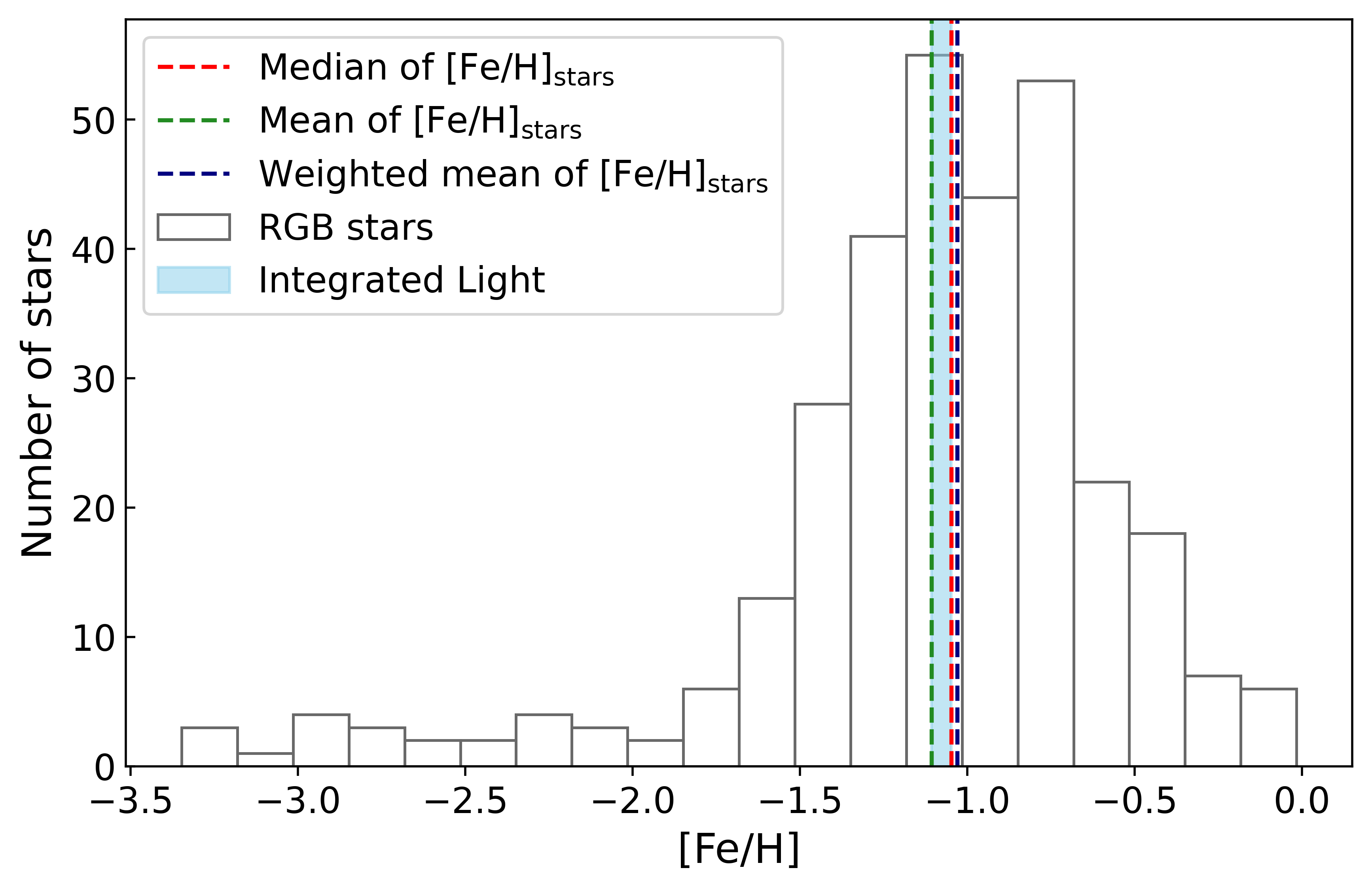}
    \caption{The metallicity distribution for the resolved stars analyzed in this sample. The value of [Fe/H]$_{\rm SSP}$ (light blue region) is consistent with the mean (green dashed line), the median (red dashed line) and the inverse-variance weighted mean (blue dashed line) of [Fe/H]$_{\rm stars}$.}
    \label{fig:MDF_IL}
\end{figure}

Before comparing the metallicity determined from integrated light ([Fe/H]$_{\rm SSP}$) to that of the resolved stars ([Fe/H]$_{\rm stars}$), it is important to understand whether the method of averaging [Fe/H]$_{\rm stars}$ would affect the final results. The median, mean, and the inverse variance-weighted mean of [Fe/H]$_{\rm stars}$ are $-1.047\pm  0.044$, $-1.106\pm 0.031$ and $-0.983 \pm 0.024$, respectively. As shown in Fig.~\ref{fig:MDF_IL}, [Fe/H]$_{\rm SSP} = -1.069^{+0.029}_{-0.031}$ from integrated light is consistent with the median and the mean of the [Fe/H]$_{\rm stars}$ within 1$\sigma$. On the other hand, [Fe/H]$_{\rm SSP}$ is slightly lower than the weighted mean, although the difference is smaller than 0.1~dex.

We calculated the uncertainties of [Fe/H]$_{\rm SSP}$ according to the methods of \citetalias{Leethochawalit2019}, which assume perfect model spectra. This assumption is not true in reality. First, NGC 147 has multiple stellar populations, contrary to the model's SSP assumption. Second, even for a simple system with an SSP, the synthetic spectra are generated from the empirical stellar libraries and the theoretical isochrones, none of which is perfect. Therefore, we attribute the small  difference ($\lesssim0.1$~dex) between the two techniques to systematic error that is not included in the uncertainties of the best-fit parameters. Still, it is appropriate to conclude that the metallicity measured from the resolved stellar spectroscopy is consistent with that derived from the integrated-light spectroscopy for NGC 147. In addition, the consistency between the two methods indicates that the SSP assumption is sufficient to recover the true metallicity of a galaxy with a complicated SFH\@.  

The agreement between the integrated-light and resolved stellar metallicities also implies that the metallicity measured from the center of NGC 147 (covered by PCWI) is representative of the much larger region covered by DEIMOS\@. Although NGC 147 has a slightly negative metallicity gradient (see Section~\ref{sec:spatial}), the two methods agree despite their drastically different spatial coverage. If this holds true for more distant galaxies, the limited spatial coverage of galaxy surveys with fiber spectrographs would have less impact on the measurement of the stellar metallicity for galaxies with metallicity gradients as shallow as those of NGC 147. A large-sample study is essential to further investigating the limited-aperture effect on the stellar metallicity recovery but it is beyond the scope of this work.

\subsection{Magnesium enhancement}

Ideally, we would compare Mg abundances measured from integrated light to those measured from resolved stars.  It is possible to measure [Mg/Fe] from DEIMOS spectra of individual stars \citep[e.g.,][]{Kirby2010}.  However, the spectral range of the DEIMOS configuration employed by the archival spectra used in this paper does not reach the Mg~b triplet or other strong Mg lines.  Instead, the available lines have relatively high excitation potential, and they are consequently weak.  This makes it difficult to measure Mg in low-S/N spectra, including those of stars in galaxies as distant as NGC 147.  The result of attempting to measure [Mg/Fe] in such galaxies is a bias toward high values of [Mg/Fe]\footnote{See Figure 10 of \citet{Kirby2020} for a demonstration of this effect.} because low values of [Mg/Fe] result in the Mg lines becoming lost in the noise.

An alternative way to validate the [Mg/Fe]$_{\rm SSP}$ is to use the MZR for [Mg/H]. The low-mass MZR for [Mg/H] can be deduced from resolved stellar spectroscopy of MW satellite galaxies \citep{Kirby2010}, which are near enough to allow S/N that overcome the bias against low values of Mg abundances.  The high-mass MZR can be measured from integrated-light spectroscopy of more distant galaxies (\citetalias{Leethochawalit2019}).  The low-mass and high-mass MZRs for [Mg/H] predict different values at the mass of NGC 147, as is the case  for [Fe/H] (see Section~\ref{subsec:mzr}). If [Mg/H]$_{\rm SSP}$ (the sum of the best-fit [Mg/Fe]$_{\rm SSP}$ and [Fe/H]$_{\rm SSP}$) falls on the low-mass MZR ([Mg/H]) for dwarf galaxies, then we have circumstantial evidence that the two methods give similar values of Mg abundance. 

We constructed the MZR for [Mg/H] from the measurements of eight MW dwarf spheroidal (dSph) satellites \citep{Kirby2010}.  We excluded more distant galaxies due to the aforementioned weak Mg lines, and we excluded most ultrafaint dwarf galaxies due to small sample sizes.  We constructed the MZR from an orthogonal, least-squares linear regression taking into account measurement uncertainties in both axes \citep{Akritas1996}.  The resulting MZR is $\langle{\rm [Mg/H]}\rangle = -3.31 + 0.32\log (M_*/M_{\sun})$.  

\begin{figure}[t]
    \includegraphics[width=0.45\textwidth]{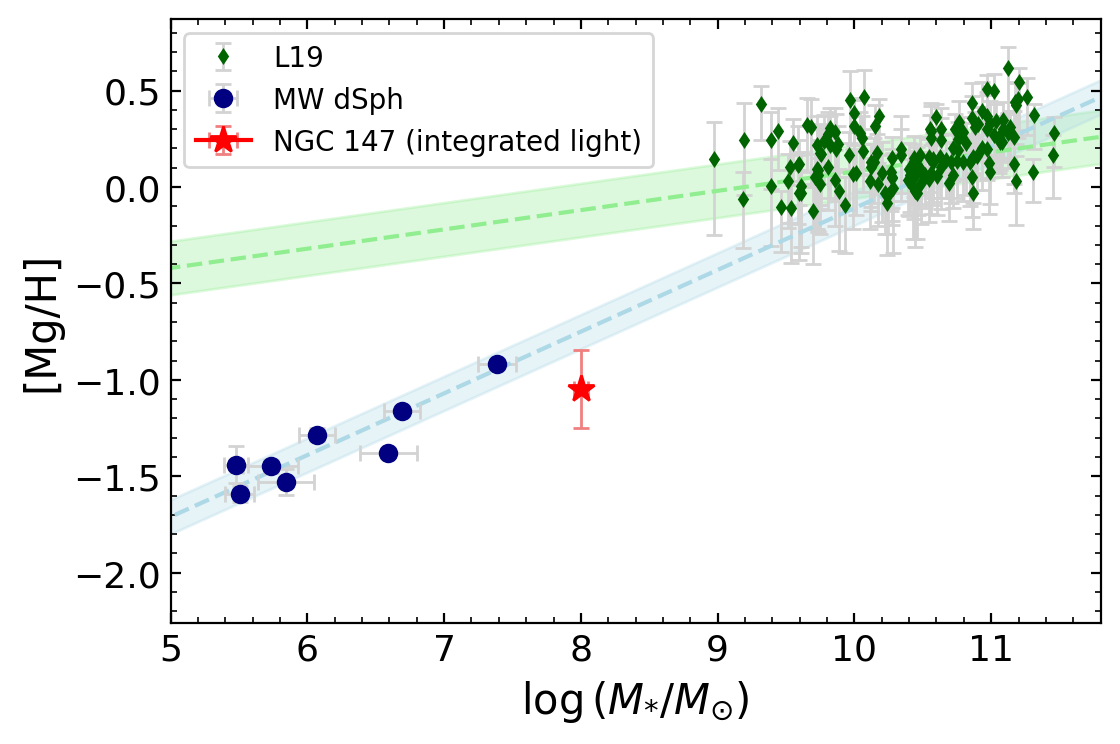}
    \caption{The MZR measured from [Mg/H] at $z\sim 0$. The green dashed line shows the high-mass MZR for Mg determined from the integrated spectra of massive quiescent galaxies (green diamonds) by \citetalias{Leethochawalit2019}. The low-mass MZR for Mg is determined only from the MW dSph satellites (blue circles). The rms for the two best-fit lines are shown in the green and blue bands. The [Mg/H] of NGC 147 measured in this work (red star) is more likely to fall on the low-mass MZR rather than on the high-mass MZR. }
    \label{fig:mzr_mgh}
\end{figure}

As seen in Fig.~\ref{fig:mzr_mgh}, the [Mg/H]$_{\rm SSP}$ of NGC 147 is much more likely to fall on the low-mass MZR (Mg) of dwarf galaxies than on the high-mass one. Considering the low-mass MZR (Mg) was determined only from a few dwarf galaxies at $M_{*} < 10^{8} M_{\odot} $, the small difference ($<$ 0.2~dex) between the [Mg/H]$_{\rm SSP}$ and the predicted value might be attributed to an extrapolation error at the mass of NGC 147 or to intrinsic dispersion in the MZR\@. We therefore conclude that we have tentative evidence that [Mg/Fe]$_{\rm SSP}$ is representative of the whole stellar population in the galaxy.

\section{Spatial Distributions of Stellar Population} \label{sec:spatial}

\subsection{Population Gradients}
\begin{figure*}[t]
    \centering
    \includegraphics[width=0.8\textwidth]{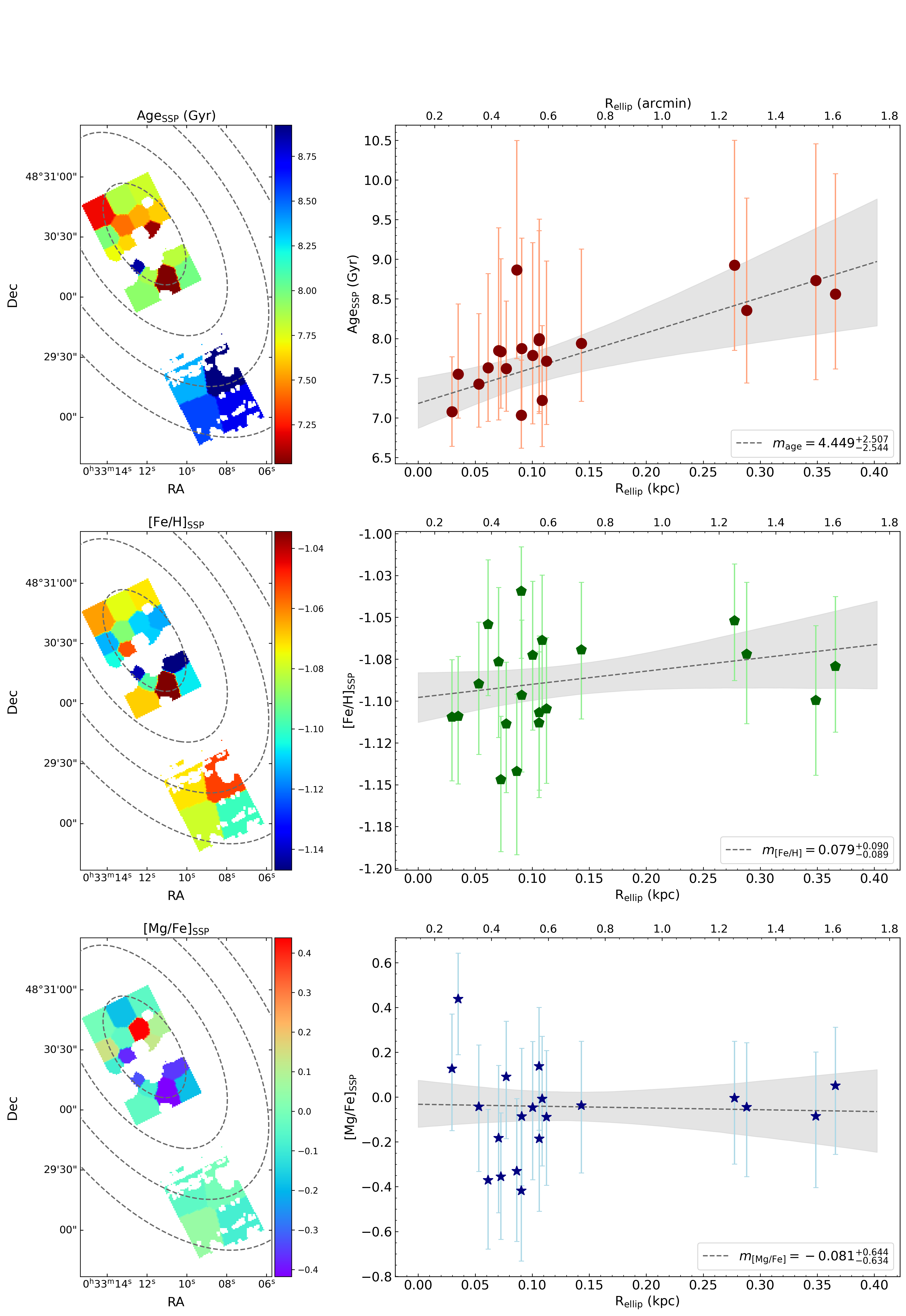}
    \caption{The spatially resolved measurements of SSP parameters for the central and south pointings. {\it Left:} The 2D maps of the best-fit stellar population age, [Fe/H], and [Mg/Fe]. The contours indicate the (projected) galactocentric elliptical radii  with 0.1~kpc  intervals.  
    {\it Right:} The radial gradients of the stellar population. The dashed lines represent the best-fit lines from MCMC. The gray shaded region indicates the ranges between the 16th and 84th percentiles.}
    \label{fig:ssp_maps}
\end{figure*}

Taking the advantage of IFS data, we were able to study the spatial distribution of the stellar population in the innermost region for the first time.  We measured the stellar population in each Voronoi cell (Section~\ref{subsubsec:spectrum_extraction}). Because the north pointing lacks reliable blue spectra below 4700 \AA, which are essential to determining the stellar population age, we excluded the north pointing for spatial distribution studies. 

Figure~\ref{fig:ssp_maps} shows the 2D maps and the radial trends for the SSP-equivalent stellar population age, [Fe/H] and [Mg/Fe]. As demonstrated by \citet{Koleva2011}, the SSP-equivalent parameters are sufficient to recover the true population gradients. We calculated the galactocentric elliptical radius using the position angle and ellipticity in \citet{Crnojevic2014}. We fit a linear model to the gradients for age, [Fe/H], and [Mg/Fe], with the \texttt{emcee} MCMC sampler \citep{Foreman-Mackey2013}. \footnote{For each gradient, the initial parameters were obtained from $\chi^2$ minimization. Broad, uniform priors were used. We ran 5000 steps with 32 walkers. Burn-in usually happened after around 40 steps, so the posteriors were well-constrained.}  The uncertainties of each parameter were taken to be the asymmetric 68\% confidence level of the posteriors. 

In the PCWI field, we marginally detected a positive age gradient within 0.4~kpc as $m_{\rm age} = \Delta \left(\frac{\rm age}{R}\right)=4.5\pm 2.5$~Gyr kpc$^{-1}$. The south field is $\sim$ 1.5~Gyr older than the innermost bin. The positive age gradient may suggest that star formation persisted longer in the central region compared to that in more extended regions. However, no age gradient is present in the \texttt{alf} measurements. As shown in Table~\ref{tab:best-fit}, around half of the best-fit results are very close to the upper limit of the stellar population age (i.e., the age of the universe), suggesting the age derived from \texttt{alf} is not well constrained. We suspect the factors discussed in Section~\ref{subsec:ssp_age} may also account for the difference. Meanwhile, the detected positive age gradient is consistent with previous literature. \citet{Han1997} also found that the younger stars in NGC 147 are more centrally concentrated than the majority of the old stars. \citet{Weisz2014} measured the SFH from {\it HST}/WFPC2 photometry in the central regions of NGC 147, which demonstrates that $\sim20\%$ of the stars were formed within the past 2.5 Gyr. On the other hand, the SFH determined from an {\it HST}/ACS field targeted at much larger radii suggests NGC 147 had formed 80\% of its stars ($6^{+2}_{-1}$)~Gyr ago  and its star formation ceased at least 3~Gyr ago \citep{Geha2015}. The different SFHs derived from HST data at different radii are qualitatively consistent with the positive age gradients we measured from integrated-light spectroscopy. 

The metallicity is relatively uniform in the PCWI fields, with a slope of $m_{\rm [Fe/H]} = \Delta \left(\frac{\rm [Fe/H]}{R}\right)=0.08\pm 0.09$~dex kpc$^{-1}$.  On the other hand, we find a large scatter for [Mg/Fe], with a slope of $m_{\rm [Mg/Fe]} = \Delta \left(\frac{\rm [Mg/Fe]}{R}\right)=-0.08^{+0.64}_{-0.63}$~dex kpc$^{-1}$. 
The large uncertainty on the [Mg/Fe] gradient precludes us from drawing conclusions about the relevance of this gradient to the galaxy's SFH\@.  Furthermore, the gradient was estimated from the very central region of the galaxy so it may not represent the whole stellar population.

\begin{figure*}[t]
    \centering
    \includegraphics[width=0.95\textwidth]{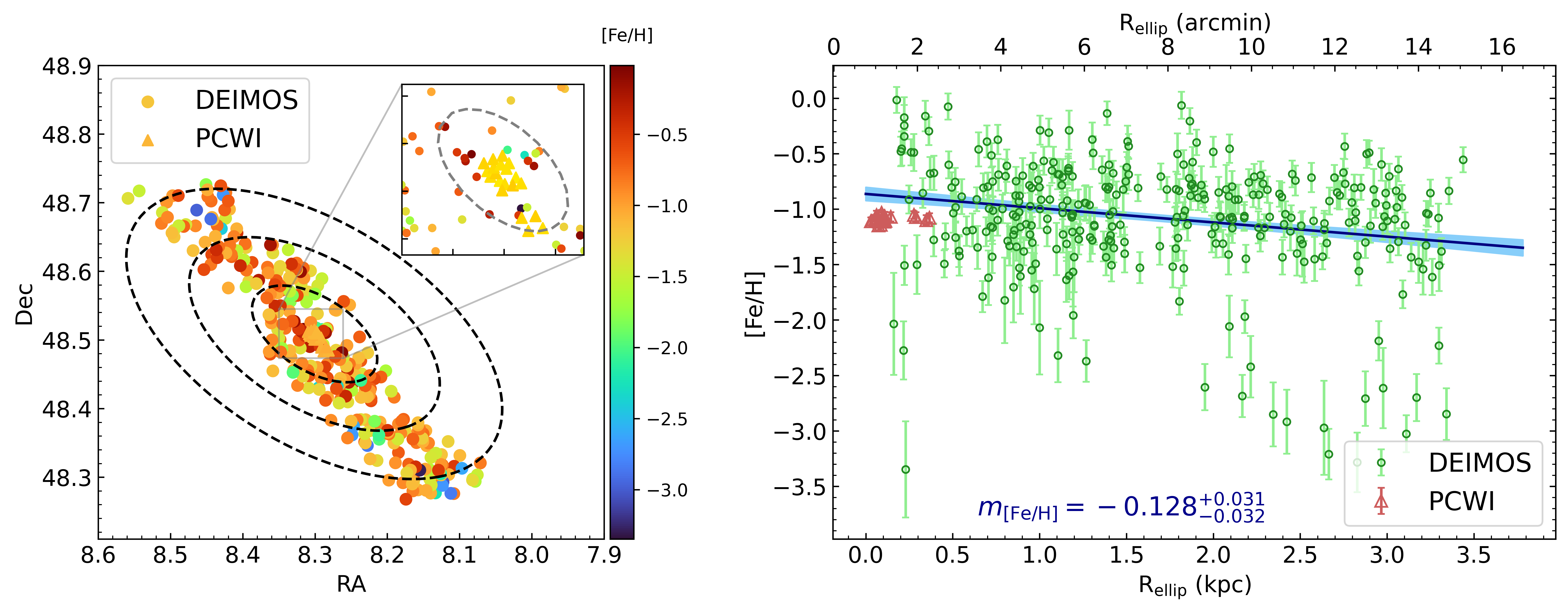}
    \caption{{\it Left panel:} The metallicity map for the resolved stars (circles) and integrated light (triangles). The black contours demonstrate the galactocentric radii at 4, 8, 12 and 16 kpc, respectively. The inset shows the zoom-in map of the galaxy central part. The gray dashed ellipse is the maximum elliptical radius of the PCWI spatial coverage. The metallicities obtained from the integrated light are very close to those of the individual RGB stars. There is no prominent spatial variance of the stars' metallicities. {\it Right panel:} The radial gradient of the metallicity. The best-fit linear gradient of the resolved stars' metallicities is shown by the purple line. A negative slope of $\sim$ -0.13~dex kpc$^{-1}$ is marginally detected.  }
    \label{fig:feh_radial_gradient}
\end{figure*}

To investigate the population gradient at larger scales, we made use of the stellar metallicities of resolved RGB stars. As shown in Figure~\ref{fig:feh_radial_gradient}, the resolved metallicities for the innermost region ($\sim$400~pc) show large scatter. There are only 17 stars within the outermost radius of the PCWI fields, so the resolved metallicities may not indicate the characteristic metallicity of the stellar population in this region. The integrated metallicities derived from PCWI data are between the lowest and highest values of the resolved metallicities. 
The resolved study reveals a negative gradient of $\Delta \frac{<{\rm [Fe/H}>}{R} \sim -0.13\pm0.03$~dex kpc$^{-1}$, different from the flat metallicity profile revealed in the PCWI fields. Because the resolved study extends to a much larger radius, we believe that the negative metallicity gradient is more representative of the galaxy population.

The negative metallicity gradient is inconsistent with the flat to positive gradient found by \citet{Crnojevic2014} and \citet{Vargas2014}. \citet{Crnojevic2014} estimated the photometric metallicity gradient from the metallicity distribution functions (MDFs) in three radial bins from 2.5 to 6.4 kpc. Their sample consists of stars at much larger radii compared to ours and lacks measurements within 2.5 kpc. In addition, the authors suggested that their gradient may be biased because stars from tidal tails can contribute to the mean stellar metallicity in the outermost region. On the other hand, \citet{Vargas2014} measured spectroscopic metallicities from Keck/DEIMOS spectra. Similar to that in \citetalias{Kirby2013}, the photometry used to estimate the effective temperature and surface gravity possibly had errors in its zero-points, leading to unreliable metallicity estimates. In fact, \citet{Vargas2014} reported a mean [Fe/H] for NGC 147 as $\sim -0.5$~dex, which is around twice the values derived in other literature, including photometric and spectroscopic estimates
\citep[][]{Han1997, Davidge2005, Goncalves2007, Geha2010, Crnojevic2014}. We recovered similarly high values of [Fe/H] when we used the questionable photometry.  We conclude that we should not compare our gradient to that measured by \citeauthor{Vargas2014}

\subsection{Outside-in Formation of NGC 147?}
Stellar population gradients can provide us insight into the evolution of a galaxy. Observational studies have suggested that dSph\footnote{Some studies call the fainter dwarf ellipticals spheroidal galaxies \citep{Kormendy2009}. }  galaxies in the Local Group generally have positive age gradients and negative metallicity gradients  \citep[e.g.,][]{Koleva2011,Kirby2011}. The observed gradients are usually explained as a result of outside-in formation\citep[e.g.,][]{Benitez-Llambay2016, Revaz2018,Genina2019}, where the metals accumulate at the center of the galaxy, and the star formation becomes more concentrated with time. In particular, \citet{Genina2019} identified that ram pressure---imposed by the massive host galaxy when the satellite galaxy falls in---can compress the gas at the center of the satellite galaxy while stripping the gas from the outskirts. Consequently, the star formation at the center can be reignited while the outer regions are quenched, giving rise to the gradients observed in the Local Group dSphs.
The age and metallicity gradients of NGC 147 found in this work are in alignment with the outside-in scenario, if the age gradient detected in the PCWI field is real.

As NGC 147 has no H$\,${\sc i} detection \citep{Young1997}, tidal stripping and ram pressure stripping are very likely to have played an important role in quenching the star formation and shaping the observed radial gradients in this galaxy. 
\citet{Crnojevic2014} uncovered the presence of extended tidal tails in NGC 147.  The tails indicate a strong tidal interaction with its host galaxy, M31.  \citet{Geha2006} found kinematic evidence for tidal disruption in another M31 dE satellite, NGC 205. \citet{Mayer2006} showed that the combined effects of tides and ram pressure are significantly more successful at removing the gas from the dwarfs than tidal stripping alone. In addition, \citet{Kormendy1985} and \citet{Kormendy2012} also suggested that ram-pressure stripping would be one of the major quenching mechanisms for dSph galaxies in the Local Group.

\section{Discussion}\label{sec:discussion}
\subsection{Comparison with previous work}\label{subsec:comparison}
So far, most works investigating the validity of recovering the stellar population, especially the metallicity, from integrated-light spectroscopy have studied relatively simple systems like globular clusters \citep[e.g.,][]{GonzaezDelgado2010,Barber2014,Conroy2018}, which tend to have formed stars at the same epoch. Very few attempts have been made to directly compare the stellar populations measured from integrated and resolved spectroscopic analyses for complex systems with multiple stellar populations. 

\citet{Ruiz-Lara2018} obtained an integrated spectrum of the central part of Leo A, one of the the Local Group's star-forming dIrr galaxies with $M_*\sim3\times10^{6}M_\odot$. They derived the SFH and the metallicity of each age bin from the integrated spectrum. They found that while the SFH is consistent with that derived from CMD fitting, the derived metallicity at all age bins is higher than the average value of individual stars obtained by \citet{Kirby2017}. The differences in metallicity range from $\sim0.25$~dex in the intermediate-age bins to $>1.0$~dex in younger-age bins, in contrast to our results for NGC 147.

However, such a discrepancy can be explained by factors other than the intrinsic systematic effects between the two techniques. First, Leo A is still star-forming and therefore has nebular emission. If the modeling of the Balmer emission lines from the ISM was not perfect, it could have affected the depths of the Balmer absorption lines, which heavily affect the measurement of the stellar population age. In turn, this could have affected the measured metallicity according to the age–metallicity degeneracy. As an old, quiescent galaxy, NGC 147 would not suffer the same source of uncertainty. Second, Leo A is metal poor ([Fe/H] $\sim$ $-$1.5). Most stellar libraries used in the modeling of integrated-light spectra do not contain enough stars with [Fe/H] $<$ $-$1. If the incompleteness of the metal-poor stars in the stellar libraries causes the discrepancy found by \citet{Ruiz-Lara2018}, it will not affect the existing metallcities of more massive and more metal-rich galaxies like NGC 147 derived from integrated light. 
In addition, Leo A exhibits a stronger metallicity gradient \citep[$\sim$ -0.33~dex/kpc; ][]{Kirby2017} than NGC 147, which may explain the discrepancy between \citet{Kirby2017} and \citet{Ruiz-Lara2018}.  \citet{Kirby2017} measured stars at larger radii ($\sim$ 7\arcmin), whereas \citet{Ruiz-Lara2018} observed a total area of 19.2\arcsec $\times$ 7\arcsec\ in the central region, corresponding to a region within a 0.15 half-light radius. 

On the other hand, \citet{Boecker2020} recovered an age-metallicity distribution of M54 measured from integrated-light spectra that was consistent with the resolved stellar spectra obtained on MUSE\@. Unlike most other star clusters, M54 has known multiple stellar populations formed at different epochs \citep{Siegel2007,Alfaro-Cuello2019}. \citeauthor{Boecker2020}\ fit for a linear combination of SSP models to account for the SFH rather than assuming an SSP, and they found excellent agreement between the mass-weighted age and metallicity. The mass-weighted metallicity of the integrated-light spectra is 0.2~dex lower than the mean for the resolved stars, whereas the difference in the light-weighted metallicity is as small as 0.1~dex. The results are similar to ours, suggesting that taking the SFH into account or not does not significantly affect the recovery of metallicity from the integrated spectra. However, their data set is slightly different from ours. First, a number of stars are resolved in the MUSE field of M54, so that the resolved stars would contribute greatly to the integrated-light spectra. Since the resolved stars are the brightest ones among the whole stellar population, one is more likely to obtain consistent light-weighted values from the integrated and resolved analyses even though systematic offsets are present. Second, both the integrated and resolved spectra are obtained on the same instrument; therefore, the instrumental effects would be minimized. Our work moves one step toward validating the comparability of integrated-light spectra of distant galaxies to resolved spectra of local galaxies because we directly compare the measurements from completely unresolved and completely resolved populations.

\subsection{Implications for the MZR}\label{subsec:mzr}
Measuring the MZR across the full range of galaxy mass is extremely important to unveil how the physical processes regulating metal retention and interaction with the surrounding environment vary as the galaxies grow in mass. However, possible systematic effects between different methods of determining the metallicity obfuscate the exact form of the MZR. 

Constructing the MZR from gas-phase metallicities has taught us an important lesson on understanding the role of systematic effects when measuring metallicities in different galaxy types and ages.  Previous studies have found that different methods used to measure gas metallicity suffer from systematic offsets \citep[e.g.,][]{Nagao2006,Kewley2008, Andrews2013,Yates2020}, and the discrepancy can be as large as 0.7~dex \citep{Kewley2008}. These systematic effects can lead to quantitatively and qualitatively different shapes for the gas-phase MZR\@. For instance, most studies favor a continuous gas-phase MZR that exhibits a single power law at low and intermediate mass, and gradually flattens at the high-mass end \citep[e.g.,][]{Tremonti2004, Kewley2008, Zahid2013}, while \citet{Blanc2019} discovered a sharp transition in the gas-phase MZR, which has a much steeper slope at $10^{9.5}-10^{10.5} M_{\odot}$. Therefore, it is extremely important to understand the systematic offsets between different methods before placing the MZRs on the same absolute scale.

The literature discussion of systematic effects on galactic stellar metallicities is small relative to the extensive studies on the discrepancies of gas metallicity diagnostics. The different treatment exists partly because it is difficult to apply resolved stellar spectroscopy and integrated-light spectroscopy to the same galaxies. It is impossible to resolve  RGB stars in massive galaxies beyond Andromeda, while most local dwarf galaxies are too large to fit in a long-slit spectrograph for integrated-light spectroscopy. So far, only \citet{Ruiz-Lara2018} have compared the stellar metallicity of Leo A determined from integrated light to that obtained from resolved stars. As discussed in Section \ref{subsec:comparison}, the large discrepancy may result from factors other than systematic offsets. 
The agreement we obtained between resolved stars and integrated light in NGC 147 is encouraging because we adopted the same fitting algorithm used to derive the high-mass MZR \citepalias{Leethochawalit2019}, which indicates relatively small systematic offsets in the stellar metallicity determination. 

\begin{figure}[t]
    \centering
    \includegraphics[width=0.45\textwidth]{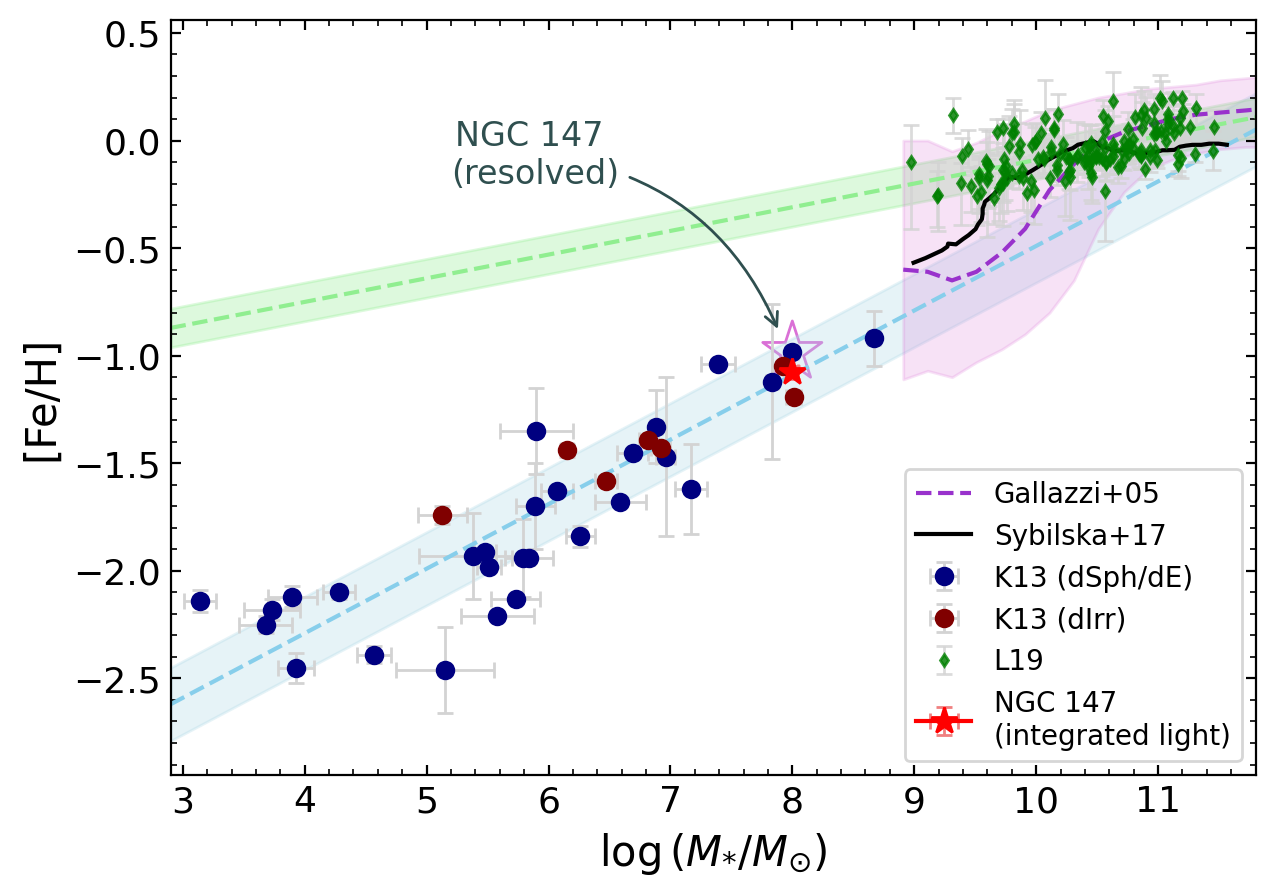}
    \caption{The MZR measured from [Fe/H] at $z\sim 0$. This figure is the same as Fig.~\ref{fig:mzr_mgh}, except that the low-mass MZR is obtained from dIrrs in addition to dSphs/dEs. The purple region shows the MZR of local Sloan Digital Sky Survey (SDSS) galaxies using the spectrophotometric indices \citep{Gallazzi2005}. The weighted mean of the [Fe/H] of NGC 147 for resolved stars, which is highlighted by the purple star, is remeasured in this work using updated photometry (Section \ref{subsec:Deimos_sample}).
    The [Fe/H] of NGC 147 measured from integrated light (red star) falls on the low-mass MZR, indicating the systematic difference between the two techniques cannot explain the discrepancy in the MZRs between the two mass ranges.}
    \label{fig:mzr_feh}
\end{figure}

As seen in Figure~\ref{fig:mzr_feh}, the high-mass MZR predicts an [Fe/H] at the mass of NGC 147 that is higher by 0.6~dex relative to the low-mass MZR\@. Our consistent resolved and integrated metallicity estimates tentatively suggest that the discrepancy between the low-mass MZR and high-mass MZR cannot be attributed to systematic effects intrinsic to the two techniques. We also exclude the possibility that the consistency in the two approaches is biased due to the SSP assumption in Appendix~\ref{appendix: comparison}.Instead, our results support a real break related to physical processes governing the metal content.
If the systematic offsets of the resolved and integrated-light methods are indeed as low as what we obtained for NGC 147, it is possible that the stellar MZR may also have a sharp transition, like the gas-phase MZR derived by \citet{Blanc2019}.
This break in the stellar MZR was also observed by \citet{Gallazzi2005}, who measured the stellar MZR from a single method (spectrophotometric indices) rather than comparing the MZRs measured from different methods in different mass ranges. However, the uncertainties of [Fe/H] below $10^{10} M_{\odot}$ are too large to distinguish between the low-mass and high-mass MZRs. 

A break in the stellar MZR was also reported by \citet{Panter2008}, who measured the stellar MZR for a large sample of local SDSS galaxies above $10^{8} M_{\odot}$ via full-spectrum fitting and discovered a steeper slope for galaxies between $10^{8}-10^{10} M_{\odot}$ than observed at higher mass. They suggested that the steeper slope was an artificial bias introduced by the limited aperture of the SDSS fibers, which cannot observe the outer region of low-mass galaxies. However, \citet{Sybilska2017} measured the stellar metallicities for dwarf and giant quiescent galaxies from IFS data cubes obtained with the SAURON IFU, which suffers much less from aperture bias. A break was still present in their MZR in the same mass range, further suggesting that the break is very likely to be real. 

The two MZRs may indicate a change in the feedback mechanisms as the galaxies grow in mass. \citet{Hayward2017} proposed an analytic model to explain how stellar feedback simultaneously regulates star formation and drives outflows in a turbulent ISM\@. They argued that the mass-loading factor could decrease dramatically for galaxies above $\sim 10^{10} M_{\odot}$, so that high-mass galaxies become much less efficient in expelling the metal-enriched ISM\@. Because the stellar metallicity represents the metals locked inside the stars from the surrounding ISM averaged over the SFH, the feedback changes can also affect the shape of the stellar MZR indirectly. Unfortunately, the current stellar MZR (Fig.~\ref{fig:mzr_feh}) has few individual precise measurements for intermediate-mass galaxies ($10^{8} < M_{\ast}/M_{\odot} < 10^{10}$), so we cannot determine how the shape of the stellar MZR would change across the full range of galaxy mass. Further studies focusing on the MZR of individual galaxies at intermediate mass are necessary to unveil the nature of the processes regulating the metal cycle and shaping the MZR\@. Nevertheless, the stellar MZRs of individual dwarf and massive galaxies can be a useful tool to help us understand the physical processes related to galactic chemical evolution. Our work presents the first evidence that it is reasonable to put the two MZRs on the same absolute scale to jointly constrain the shape of the stellar MZR.

Our investigation of systematic effects focused on just one galaxy: NGC 147. It is possible that our result is fortuitous and that a larger ensemble of galaxies would reveal some systematic differences between resolved spectroscopy and integrated light. Nonetheless, our results for NGC 147 suggest that these systematic effects do not play an important role in causing the discrepancy observed between different mass ranges of the stellar MZR. We anticipate observing more galaxies in the future to retire once and for all the possibility that the difference in technique---rather than in astrophysics---is responsible for the discrepancy. 

\section{Conclusions} \label{sec:summary}
We measured the stellar population age, [Fe/H], and [Mg/Fe] of the central region ($\lesssim$400~pc, i.e., $2'$) in NGC 147 from integrated-light spectra obtained with PCWI\@. We also provided updated metallicity estimates for 317 RGB stars in the sample of \citetalias{Kirby2013}. We compared the mean metallicity of the resolved stars with the measurement from integrated light to investigate systematic effects between the two techniques. A summary of our findings is as follows:
\begin{enumerate}
    \item  We recovered the global SSP-equivalent stellar population from integrated spectra and measured the stellar population age$=8.11^{+0.74}_{-0.54}$~Gyr, [Fe/H]$=-1.069^{+0.029}_{-0.031}$ and [Mg/Fe]$=0.024^{+0.20}_{-0.20}$. The iron and magnesium abundances are consistent with the best-fit parameters obtained from \texttt{alf}'s simple mode.

    \item The best-fit SSP age is only 0.03~dex higher than the mass-weighted age determined from the CMD fitting of \citet{Weisz2014}. The difference between [Fe/H]$_{\rm SSP}$ and the weighted mean of [Fe/H]$_{\rm stars}$ is less than 0.1~dex, implying that the SSP assumption is valid to recover the mean stellar metallicity for a complex system with multiple stellar populations. It also suggests that the SSP assumption may recover the stellar age in some cases, although it strongly depends on the choice of SSP templates. 
    
    \item Making use of the IFS cubes, we detected a marginally positive age gradient in the central regions. Combining this age gradient with the negative metallicity gradient in the resolved stars, it is possible that NGC 147 formed outside in.

    \item The resolved and integrated metallicities of NGC 147 are consistent within 0.1~dex, suggesting that systematic effects cannot explain the $\sim0.6$~dex discrepancy between the low-mass (\citetalias{Kirby2013}) and high-mass (\citetalias{Leethochawalit2019}) stellar MZRs. Therefore, we tentatively conclude that the break in the MZRs at different mass ranges implies some physical mechanism transitions at $10^{8} M_{\odot} < M_* < 10^{10} M_{\odot}$.

\end{enumerate}

There is only one target in this work, so it is possible that such a small systematic offset ($<0.1$~dex) is underestimated. Nevertheless, our analysis of NGC 147 is strong evidence that the systematic effects of different techniques of determining metallicity will not be the major cause of the discrepancy in the stellar MZRs. Future observations are necessary to expand this study to a larger sample in order to verify that the underlying astrophysical processes are ultimately responsible for the changing behavior of stellar MZR\@.


\appendix

\section{Comparison with CSP models}\label{appendix: comparison}

\begin{figure*}[t]
    \centering
    \includegraphics[width=0.85\textwidth]{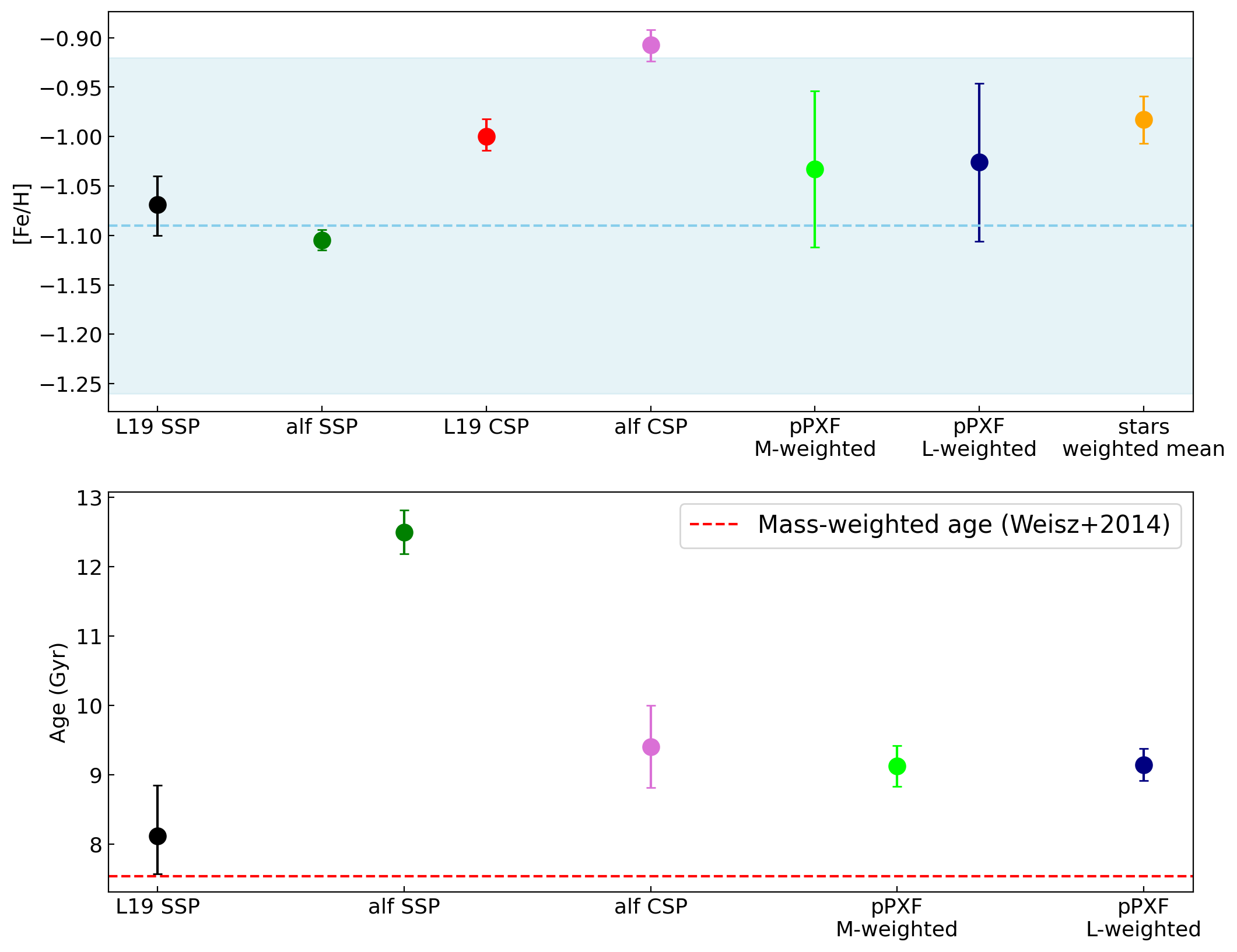}
    \caption{The comparison between the measured [Fe/H] (top panel) and stellar ages (bottom panel) in different stellar population models. {\it Top:} The blue dashed line and the shaded region indicate the [Fe/H] predicted by the low-mass MZR \citepalias{Kirby2013} at the mass of NGC 147 and the corresponding 1$\sigma$ confidence interval. The data points show the measured [Fe/H] (weighted mean for CSP models) via various methods discussed in this work, which are all consistent with the low-mass MZR\@.  {\it Bottom:} the measured ages (weighted mean for CSP models) via different methods discussed in this work. Among them, only the SSP age derived from L19's method is consistent with the mass-weighted age obtained from the CMD fitting in \citet{Weisz2014} (red dashed line).}
    \label{fig:model_comparison}
\end{figure*}

In this section, we investigate whether fitting the data with more complicated CSP models would affect our primary results. Because CSP models have more free parameters and thus require higher-quality data to achieve the same precision, we limit our analysis to the stacked spectrum, which has very high S/N ($\sim111$\ \AA$^{-1}$). Here we present the measured ages and [Fe/H] from three different full spectral fitting codes that fit for multicomponent stellar populations: \citetalias{Leethochawalit2019}'s method (Appendix~\ref{appendix:L19}), \texttt{alf} (Appendix~\ref{appendix:alf}), and \textsc{pPXF} \citep[][Appendix~\ref{appendix:ppxf}]{Cappellari2017}. As shown in Figure~\ref{fig:model_comparison}, the stellar metallicity estimates are not greatly affected by the choice of models, while the measured stellar ages are more strongly dependent on the assumed SFH\@. We find that taking the SFH into account leads to greater consistency between different methods and templates than assuming an SSP\@. Therefore, CSP models may be a more appropriate way to estimate the stellar age of complex systems when the data quality is good enough.

\subsection{Comparison to the CSP models in \citetalias{Leethochawalit2019}'s fitting code}\label{appendix:L19}

To investigate whether adopting CSP models would affect the recovered metallicities, we fit for a multicomponent stellar population with ages and weights determined from the SFH measured from CMD fitting to {\it HST} photometry \citep{Weisz2014}. Instead of fitting for linear combinations of SSP models with different metallicities, we choose to unrealistically assume that all components have the same metallicity. This can reduce the degeneracy of free parameters and approximately estimate the mass-weighted metallicities to the greatest accuracy. 

The best-fit metallicities are [Fe/H]$_{\rm CSP}$ =$-1.000^{+0.018}_{-0.014}$ and [Mg/Fe]$_{\rm CSP}$ =$0.096^{+0.15}_{-0.37}$. Compared to [Fe/H]$_{\rm SSP}=-1.069^{+0.029}_{-0.031}$ and [Mg/Fe]$_{\rm SSP}=0.024^{+0.20}_{-0.20}$, the best-fit metallicities derived from CSP models are higher and closer to the resolved metallicities, further diminishing the systematic offsets between the two techniques. The differences in the recovered [Fe/H] and [Mg/Fe] from SSP models and CSP models are within 0.1~dex; therefore, we conclude that the metallicity measurements are not affected significantly by the choices of SSP and CSP models, consistent with the results of \citet{Mentz2016} and \citet{Leethochawalit2018}.

\subsection{Comparison to the results from \texttt{alf} full mode}\label{appendix:alf}
In addition to the SSP models in simple mode, \texttt{alf} also provides a full mode that fits for two-component stellar populations with an older ($>$3~Gyr) and a younger population ($<$3~Gyr), which allows us to approximate the SFH of the galaxy. We fit the stacked spectrum in \texttt{alf} full mode. 

    The mass-weighted age derived in full mode (CSP-equivalent) is $\sim$ 9.4~Gyr. Compared to the SSP-equivalent age (12.5~Gyr) derived in simple mode, the difference between the CSP age and the mass-weighted age from CMD fitting by \citet{Weisz2014} is smaller. The consistency suggests that more complicated stellar population models are necessary to obtain more accurate stellar age measurements in galaxies. On the other hand, the best-fit results from \texttt{alf} full mode indicate that less than 0.5\% of the stellar population was formed in the past 3~Gyr, while the SFH revealed by the CMD fitting in \citet{Weisz2014} suggests that $\sim$20\% of stars were formed in the past 2.5~Gyr. We also notice that the CSP age is surprisingly younger than the SSP age, whereas \citet{Choi2014} found that SSP-equivalent ages are younger than both light- and mass-weighted ages derived in models with extended SFHs. The discrepancy may arise from the fact that \texttt{alf} assumes two independent starbursts at different epochs rather than extended SFHs. The disagreements above may also suggest that the recovered stellar age is sensitive to the adopted CSP models and their underlying assumptions on SFH, and thus future investigations are important to explore which CSP models are more suitable to measuring the stellar ages of galaxies.

Unlike the large difference seen in ages, the metallicities derived in the two modes are very similar. [Mg/Fe]$_{\rm CSP}$ =$0.064^{+0.018}_{-0.016}$) is consistent with [Mg/Fe]$_{\rm SSP}$ within $1\sigma$, while [Fe/H]$_{\rm CSP}$ = $-0.907^{+0.015}_{-0.017}$ is slightly higher than [Fe/H]$_{\rm SSP}$. The difference between [Fe/H]$_{\rm CSP}$ and [Fe/H]$_{\rm stars}$ is still within 0.1~dex, which once again demonstrates that the metallicity measurements are less affected by the assumed stellar population models.

\subsection{Comparisons to \textsc{pPXF} full spectral fitting code}\label{appendix:ppxf}
Apart from the two fitting methods discussed in the main text, we test our CSP measurements by running another independent code, \textsc{pPXF} \citep{Cappellari2017}, which fits for linear combinations of different SSPs to account for complex SFHs in galaxies. For simplicity, $\alpha$-enhancement is not included. We use the SSP models from the E-MILES library \citep{Vazdekis2016}, generated from the BaSTI isochrones \citep{Pietrinferni2004} and Kroupa \citep{Kroupa2001} IMF\@. The metallicities and ages of the templates range from $-2.27$ to 0.4~dex and 0.03 to 14~Gyr, respectively. Because the SSP models assume solar-scaled abundances, the metallicity [M/H] can be interpreted as [Fe/H]. Regularization is used to address some of the deficiencies in \textsc{pPXF}: that it does not necessarily return unique solutions, and that it is very sensitive to the noise in data due to the vast parameter space explored in fitting. 
The mass- and light-weighted stellar population parameters can be computed from the weights of each SSP component returned by \textsc{pPXF}. 

The mass- and light-weighted [Fe/H] are $-1.033\pm 0.079$ and $-1.026\pm 0.08$, respectively, consistent with the metallicities measured from other methods. The uncertainties are calculated as the weighted variance of the ages and metallicities of each SSP component. On the other hand, the mass- and light-weighted ages are both $9.96\pm 0.01$~Gyr, higher than the SFH-based age measured by \citet{Weisz2014} but similar to the mass-weighted age derived in \texttt{alf} full mode, which also uses CSP models. The results suggest that stellar age estimates from integrated-light spectroscopy depend more heavily than metallicity estimates on the assumed SFH\@.

\begin{acknowledgments}
The authors acknowledge the constructive feedback from the anonymous referee, which helped us improve the manuscript. We thank Yuguang Chen, Donal O'Sullivan and James (Don) Neill for useful discussions on PCWI data reduction and covariance correction. We gratefully thank the staff at the Palomar Observatory, including support astronomers Carolyn Heffner and Kevin Rykoski, and telescope operator Kajsa Peffer, for assisting in the observations.

This material is based on work supported by the National Science Foundation under Grant No.\ AST-1847909.  E.N.K.\ gratefully acknowledges support from a Cottrell Scholar Award administered by the Research Corporation for Science Advancement.  M.A.dl.R. acknowledges the financial support of the NSF Graduate Research Fellowship Program.

We are grateful to the many people who have worked to make the Keck Telescope and its instruments a reality and to operate and maintain the Keck Observatory.  The authors wish to extend special thanks to those of Hawaiian ancestry, on whose sacred mountain we are privileged to be guests.  Without their generous hospitality, none of the observations presented herein would have been possible.  This research has made use of the Keck Observatory Archive, which is operated by the W. M. Keck Observatory and the NASA Exoplanet Science Institute, under a contract with the National Aeronautics and Space Administration. 

\facilities{Hale (CWI), Keck II	(DEIMOS)}
\software{CWITools \citep{OSullivan2020b}, FSPS \citep{Conroy2009}, alf \citep{Conroy2018}, MPFIT \citep{Markwardt2012}, Astropy \citep{Astropy2013,Astropy2018}, NumPy \citep{2020NumPy-Array}, SciPy \citep{2020SciPy-NMeth}}, pPXF \citep{Cappellari2017}

\end{acknowledgments}

\bibliography{main}

\begin{thebibliography}{}
\expandafter\ifx\csname natexlab\endcsname\relax\def\natexlab#1{#1}\fi
\providecommand{\url}[1]{\href{#1}{#1}}
\providecommand{\dodoi}[1]{doi:~\href{http://doi.org/#1}{\nolinkurl{#1}}}
\providecommand{\doeprint}[1]{\href{http://ascl.net/#1}{\nolinkurl{http://ascl.net/#1}}}
\providecommand{\doarXiv}[1]{\href{https://arxiv.org/abs/#1}{\nolinkurl{https://arxiv.org/abs/#1}}}

\bibitem[{{Akritas} \& {Bershady}(1996)}]{Akritas1996}
{Akritas}, M.~G., \& {Bershady}, M.~A. 1996, \apj, 470, 706,
  \dodoi{10.1086/177901}

\bibitem[{{Alfaro-Cuello} {et~al.}(2019){Alfaro-Cuello}, {Kacharov},
  {Neumayer}, {L{\"u}tzgendorf}, {Seth}, {B{\"o}ker}, {Kamann}, {Leaman}, {van
  de Ven}, {Bianchini}, {Watkins}, \& {Lyubenova}}]{Alfaro-Cuello2019}
{Alfaro-Cuello}, M., {Kacharov}, N., {Neumayer}, N., {et~al.} 2019, \apj, 886,
  57, \dodoi{10.3847/1538-4357/ab1b2c}

\bibitem[{{Andrews} \& {Martini}(2013)}]{Andrews2013}
{Andrews}, B.~H., \& {Martini}, P. 2013, \apj, 765, 140,
  \dodoi{10.1088/0004-637X/765/2/140}

\bibitem[{{Astropy Collaboration} {et~al.}(2013){Astropy Collaboration},
  {Robitaille}, {Tollerud}, {Greenfield}, {Droettboom}, {Bray}, {Aldcroft},
  {Davis}, {Ginsburg}, {Price-Whelan}, {Kerzendorf}, {Conley}, {Crighton},
  {Barbary}, {Muna}, {Ferguson}, {Grollier}, {Parikh}, {Nair}, {Unther},
  {Deil}, {Woillez}, {Conseil}, {Kramer}, {Turner}, {Singer}, {Fox}, {Weaver},
  {Zabalza}, {Edwards}, {Azalee Bostroem}, {Burke}, {Casey}, {Crawford},
  {Dencheva}, {Ely}, {Jenness}, {Labrie}, {Lim}, {Pierfederici}, {Pontzen},
  {Ptak}, {Refsdal}, {Servillat}, \& {Streicher}}]{Astropy2013}
{Astropy Collaboration}, {Robitaille}, T.~P., {Tollerud}, E.~J., {et~al.} 2013,
  \aap, 558, A33, \dodoi{10.1051/0004-6361/201322068}

\bibitem[{{Astropy Collaboration} {et~al.}(2018){Astropy Collaboration},
  {Price-Whelan}, {Sip{\H{o}}cz}, {G{\"u}nther}, {Lim}, {Crawford}, {Conseil},
  {Shupe}, {Craig}, {Dencheva}, {Ginsburg}, {VanderPlas}, {Bradley},
  {P{\'e}rez-Su{\'a}rez}, {de Val-Borro}, {Aldcroft}, {Cruz}, {Robitaille},
  {Tollerud}, {Ardelean}, {Babej}, {Bach}, {Bachetti}, {Bakanov}, {Bamford},
  {Barentsen}, {Barmby}, {Baumbach}, {Berry}, {Biscani}, {Boquien}, {Bostroem},
  {Bouma}, {Brammer}, {Bray}, {Breytenbach}, {Buddelmeijer}, {Burke},
  {Calderone}, {Cano Rodr{\'\i}guez}, {Cara}, {Cardoso}, {Cheedella}, {Copin},
  {Corrales}, {Crichton}, {D'Avella}, {Deil}, {Depagne}, {Dietrich}, {Donath},
  {Droettboom}, {Earl}, {Erben}, {Fabbro}, {Ferreira}, {Finethy}, {Fox},
  {Garrison}, {Gibbons}, {Goldstein}, {Gommers}, {Greco}, {Greenfield},
  {Groener}, {Grollier}, {Hagen}, {Hirst}, {Homeier}, {Horton}, {Hosseinzadeh},
  {Hu}, {Hunkeler}, {Ivezi{\'c}}, {Jain}, {Jenness}, {Kanarek}, {Kendrew},
  {Kern}, {Kerzendorf}, {Khvalko}, {King}, {Kirkby}, {Kulkarni}, {Kumar},
  {Lee}, {Lenz}, {Littlefair}, {Ma}, {Macleod}, {Mastropietro}, {McCully},
  {Montagnac}, {Morris}, {Mueller}, {Mumford}, {Muna}, {Murphy}, {Nelson},
  {Nguyen}, {Ninan}, {N{\"o}the}, {Ogaz}, {Oh}, {Parejko}, {Parley}, {Pascual},
  {Patil}, {Patil}, {Plunkett}, {Prochaska}, {Rastogi}, {Reddy Janga},
  {Sabater}, {Sakurikar}, {Seifert}, {Sherbert}, {Sherwood-Taylor}, {Shih},
  {Sick}, {Silbiger}, {Singanamalla}, {Singer}, {Sladen}, {Sooley},
  {Sornarajah}, {Streicher}, {Teuben}, {Thomas}, {Tremblay}, {Turner},
  {Terr{\'o}n}, {van Kerkwijk}, {de la Vega}, {Watkins}, {Weaver}, {Whitmore},
  {Woillez}, {Zabalza}, \& {Astropy Contributors}}]{Astropy2018}
{Astropy Collaboration}, {Price-Whelan}, A.~M., {Sip{\H{o}}cz}, B.~M., {et~al.}
  2018, \aj, 156, 123, \dodoi{10.3847/1538-3881/aabc4f}

\bibitem[{{Barber} {et~al.}(2014){Barber}, {Courteau}, {Roediger}, \&
  {Schiavon}}]{Barber2014}
{Barber}, C., {Courteau}, S., {Roediger}, J.~C., \& {Schiavon}, R.~P. 2014,
  \mnras, 440, 2953, \dodoi{10.1093/mnras/stu439}

\bibitem[{{Ben{\'\i}tez-Llambay} {et~al.}(2016){Ben{\'\i}tez-Llambay},
  {Navarro}, {Abadi}, {Gottl{\"o}ber}, {Yepes}, {Hoffman}, \&
  {Steinmetz}}]{Benitez-Llambay2016}
{Ben{\'\i}tez-Llambay}, A., {Navarro}, J.~F., {Abadi}, M.~G., {et~al.} 2016,
  \mnras, 456, 1185, \dodoi{10.1093/mnras/stv2722}

\bibitem[{{Blanc} {et~al.}(2019){Blanc}, {Lu}, {Benson}, {Katsianis}, \&
  {Barraza}}]{Blanc2019}
{Blanc}, G.~A., {Lu}, Y., {Benson}, A., {Katsianis}, A., \& {Barraza}, M. 2019,
  \apj, 877, 6, \dodoi{10.3847/1538-4357/ab16ec}

\bibitem[{{Boecker} {et~al.}(2020){Boecker}, {Alfaro-Cuello}, {Neumayer},
  {Mart{\'\i}n-Navarro}, \& {Leaman}}]{Boecker2020}
{Boecker}, A., {Alfaro-Cuello}, M., {Neumayer}, N., {Mart{\'\i}n-Navarro}, I.,
  \& {Leaman}, R. 2020, \apj, 896, 13, \dodoi{10.3847/1538-4357/ab919d}

\bibitem[{{Calura} {et~al.}(2009){Calura}, {Pipino}, {Chiappini}, {Matteucci},
  \& {Maiolino}}]{Calura2009}
{Calura}, F., {Pipino}, A., {Chiappini}, C., {Matteucci}, F., \& {Maiolino}, R.
  2009, \aap, 504, 373, \dodoi{10.1051/0004-6361/200911756}

\bibitem[{{Cappellari}(2017)}]{Cappellari2017}
{Cappellari}, M. 2017, \mnras, 466, 798, \dodoi{10.1093/mnras/stw3020}

\bibitem[{{Cappellari} \& {Copin}(2003)}]{Cappellari2003}
{Cappellari}, M., \& {Copin}, Y. 2003, \mnras, 342, 345,
  \dodoi{10.1046/j.1365-8711.2003.06541.x}

\bibitem[{{Cappellari} {et~al.}(2011){Cappellari}, {Emsellem}, {Krajnovi{\'c}},
  {McDermid}, {Scott}, {Verdoes Kleijn}, {Young}, {Alatalo}, {Bacon}, {Blitz},
  {Bois}, {Bournaud}, {Bureau}, {Davies}, {Davis}, {de Zeeuw}, {Duc},
  {Khochfar}, {Kuntschner}, {Lablanche}, {Morganti}, {Naab}, {Oosterloo},
  {Sarzi}, {Serra}, \& {Weijmans}}]{Cappellari2011}
{Cappellari}, M., {Emsellem}, E., {Krajnovi{\'c}}, D., {et~al.} 2011, \mnras,
  413, 813, \dodoi{10.1111/j.1365-2966.2010.18174.x}

\bibitem[{{Carnall}(2017)}]{Carnall2017}
{Carnall}, A.~C. 2017, arXiv e-prints, arXiv:1705.05165.
\newblock \doarXiv{1705.05165}

\bibitem[{{Choi} {et~al.}(2014){Choi}, {Conroy}, {Moustakas}, {Graves},
  {Holden}, {Brodwin}, {Brown}, \& {van Dokkum}}]{Choi2014}
{Choi}, J., {Conroy}, C., {Moustakas}, J., {et~al.} 2014, \apj, 792, 95,
  \dodoi{10.1088/0004-637X/792/2/95}

\bibitem[{{Choi} {et~al.}(2016){Choi}, {Dotter}, {Conroy}, {Cantiello},
  {Paxton}, \& {Johnson}}]{Choi2016}
{Choi}, J., {Dotter}, A., {Conroy}, C., {et~al.} 2016, \apj, 823, 102,
  \dodoi{10.3847/0004-637X/823/2/102}

\bibitem[{{Conn} {et~al.}(2012){Conn}, {Ibata}, {Lewis}, {Parker}, {Zucker},
  {Martin}, {McConnachie}, {Irwin}, {Tanvir}, {Fardal}, {Ferguson}, {Chapman},
  \& {Valls-Gabaud}}]{Conn2012}
{Conn}, A.~R., {Ibata}, R.~A., {Lewis}, G.~F., {et~al.} 2012, \apj, 758, 11,
  \dodoi{10.1088/0004-637X/758/1/11}

\bibitem[{{Conroy} {et~al.}(2014){Conroy}, {Graves}, \& {van
  Dokkum}}]{Conroy2014}
{Conroy}, C., {Graves}, G.~J., \& {van Dokkum}, P.~G. 2014, \apj, 780, 33,
  \dodoi{10.1088/0004-637X/780/1/33}

\bibitem[{{Conroy} {et~al.}(2009){Conroy}, {Gunn}, \& {White}}]{Conroy2009}
{Conroy}, C., {Gunn}, J.~E., \& {White}, M. 2009, \apj, 699, 486,
  \dodoi{10.1088/0004-637X/699/1/486}

\bibitem[{{Conroy} {et~al.}(2018){Conroy}, {Villaume}, {van Dokkum}, \&
  {Lind}}]{Conroy2018}
{Conroy}, C., {Villaume}, A., {van Dokkum}, P.~G., \& {Lind}, K. 2018, \apj,
  854, 139, \dodoi{10.3847/1538-4357/aaab49}

\bibitem[{{Conroy} {et~al.}(2010){Conroy}, {White}, \& {Gunn}}]{Conroy2010}
{Conroy}, C., {White}, M., \& {Gunn}, J.~E. 2010, \apj, 708, 58,
  \dodoi{10.1088/0004-637X/708/1/58}

\bibitem[{{Cooper} {et~al.}(2012){Cooper}, {Griffith}, {Newman}, {Coil},
  {Davis}, {Dutton}, {Faber}, {Guhathakurta}, {Koo}, {Lotz}, {Weiner},
  {Willmer}, \& {Yan}}]{Cooper2012}
{Cooper}, M.~C., {Griffith}, R.~L., {Newman}, J.~A., {et~al.} 2012, \mnras,
  419, 3018, \dodoi{10.1111/j.1365-2966.2011.19938.x}

\bibitem[{{Crnojevi{\'c}} {et~al.}(2014){Crnojevi{\'c}}, {Ferguson}, {Irwin},
  {McConnachie}, {Bernard}, {Fardal}, {Ibata}, {Lewis}, {Martin}, {Navarro},
  {No{\"e}l}, \& {Pasetto}}]{Crnojevic2014}
{Crnojevi{\'c}}, D., {Ferguson}, A.~M.~N., {Irwin}, M.~J., {et~al.} 2014,
  \mnras, 445, 3862, \dodoi{10.1093/mnras/stu2003}

\bibitem[{{Davidge}(2005)}]{Davidge2005}
{Davidge}, T.~J. 2005, \aj, 130, 2087, \dodoi{10.1086/491706}

\bibitem[{{Dekel} \& {Silk}(1986)}]{Dekel1986}
{Dekel}, A., \& {Silk}, J. 1986, \apj, 303, 39, \dodoi{10.1086/164050}

\bibitem[{{Dotter}(2016)}]{Dotter2016}
{Dotter}, A. 2016, \apjs, 222, 8, \dodoi{10.3847/0067-0049/222/1/8}

\bibitem[{{Faber} {et~al.}(2003){Faber}, {Phillips}, {Kibrick}, {Alcott},
  {Allen}, {Burrous}, {Cantrall}, {Clarke}, {Coil}, {Cowley}, {Davis}, {Deich},
  {Dietsch}, {Gilmore}, {Harper}, {Hilyard}, {Lewis}, {McVeigh}, {Newman},
  {Osborne}, {Schiavon}, {Stover}, {Tucker}, {Wallace}, {Wei}, {Wirth}, \&
  {Wright}}]{Faber2003}
{Faber}, S.~M., {Phillips}, A.~C., {Kibrick}, R.~I., {et~al.} 2003, in Society
  of Photo-Optical Instrumentation Engineers (SPIE) Conference Series, Vol.
  4841, Instrument Design and Performance for Optical/Infrared Ground-based
  Telescopes, ed. M.~{Iye} \& A.~F.~M. {Moorwood}, 1657--1669,
  \dodoi{10.1117/12.460346}

\bibitem[{{Finlator} \& {Dav{\'e}}(2008)}]{Finlator2008}
{Finlator}, K., \& {Dav{\'e}}, R. 2008, \mnras, 385, 2181,
  \dodoi{10.1111/j.1365-2966.2008.12991.x}

\bibitem[{{Foreman-Mackey} {et~al.}(2013){Foreman-Mackey}, {Hogg}, {Lang}, \&
  {Goodman}}]{Foreman-Mackey2013}
{Foreman-Mackey}, D., {Hogg}, D.~W., {Lang}, D., \& {Goodman}, J. 2013, \pasp,
  125, 306, \dodoi{10.1086/670067}

\bibitem[{{Gallazzi} {et~al.}(2005){Gallazzi}, {Charlot}, {Brinchmann},
  {White}, \& {Tremonti}}]{Gallazzi2005}
{Gallazzi}, A., {Charlot}, S., {Brinchmann}, J., {White}, S. D.~M., \&
  {Tremonti}, C.~A. 2005, \mnras, 362, 41,
  \dodoi{10.1111/j.1365-2966.2005.09321.x}

\bibitem[{{Geha} {et~al.}(2006){Geha}, {Guhathakurta}, {Rich}, \&
  {Cooper}}]{Geha2006}
{Geha}, M., {Guhathakurta}, P., {Rich}, R.~M., \& {Cooper}, M.~C. 2006, \aj,
  131, 332, \dodoi{10.1086/498686}

\bibitem[{{Geha} {et~al.}(2010){Geha}, {van der Marel}, {Guhathakurta},
  {Gilbert}, {Kalirai}, \& {Kirby}}]{Geha2010}
{Geha}, M., {van der Marel}, R.~P., {Guhathakurta}, P., {et~al.} 2010, \apj,
  711, 361, \dodoi{10.1088/0004-637X/711/1/361}

\bibitem[{{Geha} {et~al.}(2015){Geha}, {Weisz}, {Grocholski}, {Dolphin}, {van
  der Marel}, \& {Guhathakurta}}]{Geha2015}
{Geha}, M., {Weisz}, D., {Grocholski}, A., {et~al.} 2015, \apj, 811, 114,
  \dodoi{10.1088/0004-637X/811/2/114}

\bibitem[{{Genina} {et~al.}(2019){Genina}, {Frenk}, {Ben{\'\i}tez-Llambay},
  {Cole}, {Navarro}, {Oman}, \& {Fattahi}}]{Genina2019}
{Genina}, A., {Frenk}, C.~S., {Ben{\'\i}tez-Llambay}, A., {et~al.} 2019,
  \mnras, 488, 2312, \dodoi{10.1093/mnras/stz1852}

\bibitem[{{Gon{\c{c}}alves} {et~al.}(2007){Gon{\c{c}}alves}, {Magrini},
  {Leisy}, \& {Corradi}}]{Goncalves2007}
{Gon{\c{c}}alves}, D.~R., {Magrini}, L., {Leisy}, P., \& {Corradi}, R.~L.~M.
  2007, \mnras, 375, 715, \dodoi{10.1111/j.1365-2966.2006.11339.x}

\bibitem[{{Gonz{\'a}lez Delgado} \& {Cid Fernandes}(2010)}]{GonzaezDelgado2010}
{Gonz{\'a}lez Delgado}, R.~M., \& {Cid Fernandes}, R. 2010, \mnras, 403, 797,
  \dodoi{10.1111/j.1365-2966.2009.16152.x}

\bibitem[{{Han} {et~al.}(1997){Han}, {Hoessel}, {Gallagher}, {Holtsman},
  {Stetson}, {Trauger}, {Ballester}, {Burrows}, {Clarke}, {Crisp}, {Griffiths},
  {Grillmair}, {Hester}, {Krist}, {Mould}, {Scowen}, {Stapelfeldt}, {Watson},
  \& {Westphal}}]{Han1997}
{Han}, M., {Hoessel}, J.~G., {Gallagher}, J.~S., I., {et~al.} 1997, \aj, 113,
  1001, \dodoi{10.1086/118316}

\bibitem[{Harris {et~al.}(2020)Harris, Millman, van~der Walt, Gommers,
  Virtanen, Cournapeau, Wieser, Taylor, Berg, Smith, Kern, Picus, Hoyer, van
  Kerkwijk, Brett, Haldane, Fernández~del Río, Wiebe, Peterson,
  Gérard-Marchant, Sheppard, Reddy, Weckesser, Abbasi, Gohlke, \&
  Oliphant}]{2020NumPy-Array}
Harris, C.~R., Millman, K.~J., van~der Walt, S.~J., {et~al.} 2020, Nature, 585,
  357–362, \dodoi{10.1038/s41586-020-2649-2}

\bibitem[{{Hayward} \& {Hopkins}(2017)}]{Hayward2017}
{Hayward}, C.~C., \& {Hopkins}, P.~F. 2017, \mnras, 465, 1682,
  \dodoi{10.1093/mnras/stw2888}

\bibitem[{{Hopkins} {et~al.}(2012){Hopkins}, {Quataert}, \&
  {Murray}}]{Hopkins2012}
{Hopkins}, P.~F., {Quataert}, E., \& {Murray}, N. 2012, \mnras, 421, 3522,
  \dodoi{10.1111/j.1365-2966.2012.20593.x}

\bibitem[{{Kewley} \& {Ellison}(2008)}]{Kewley2008}
{Kewley}, L.~J., \& {Ellison}, S.~L. 2008, \apj, 681, 1183,
  \dodoi{10.1086/587500}

\bibitem[{{Kirby} {et~al.}(2013){Kirby}, {Cohen}, {Guhathakurta}, {Cheng},
  {Bullock}, \& {Gallazzi}}]{Kirby2013}
{Kirby}, E.~N., {Cohen}, J.~G., {Guhathakurta}, P., {et~al.} 2013, \apj, 779,
  102, \dodoi{10.1088/0004-637X/779/2/102}

\bibitem[{{Kirby} {et~al.}(2020){Kirby}, {Gilbert}, {Escala}, {Wojno},
  {Guhathakurta}, {Majewski}, \& {Beaton}}]{Kirby2020}
{Kirby}, E.~N., {Gilbert}, K.~M., {Escala}, I., {et~al.} 2020, \aj, 159, 46,
  \dodoi{10.3847/1538-3881/ab5f0f}

\bibitem[{{Kirby} {et~al.}(2009){Kirby}, {Guhathakurta}, {Bolte}, {Sneden}, \&
  {Geha}}]{Kirby2009}
{Kirby}, E.~N., {Guhathakurta}, P., {Bolte}, M., {Sneden}, C., \& {Geha}, M.~C.
  2009, \apj, 705, 328, \dodoi{10.1088/0004-637X/705/1/328}

\bibitem[{{Kirby} {et~al.}(2008){Kirby}, {Guhathakurta}, \&
  {Sneden}}]{Kirby2008}
{Kirby}, E.~N., {Guhathakurta}, P., \& {Sneden}, C. 2008, \apj, 682, 1217,
  \dodoi{10.1086/589627}

\bibitem[{{Kirby} {et~al.}(2011){Kirby}, {Lanfranchi}, {Simon}, {Cohen}, \&
  {Guhathakurta}}]{Kirby2011}
{Kirby}, E.~N., {Lanfranchi}, G.~A., {Simon}, J.~D., {Cohen}, J.~G., \&
  {Guhathakurta}, P. 2011, \apj, 727, 78, \dodoi{10.1088/0004-637X/727/2/78}

\bibitem[{{Kirby} {et~al.}(2017){Kirby}, {Rizzi}, {Held}, {Cohen}, {Cole},
  {Manning}, {Skillman}, \& {Weisz}}]{Kirby2017}
{Kirby}, E.~N., {Rizzi}, L., {Held}, E.~V., {et~al.} 2017, \apj, 834, 9,
  \dodoi{10.3847/1538-4357/834/1/9}

\bibitem[{{Kirby} {et~al.}(2010){Kirby}, {Guhathakurta}, {Simon}, {Geha},
  {Rockosi}, {Sneden}, {Cohen}, {Sohn}, {Majewski}, \& {Siegel}}]{Kirby2010}
{Kirby}, E.~N., {Guhathakurta}, P., {Simon}, J.~D., {et~al.} 2010, \apjs, 191,
  352, \dodoi{10.1088/0067-0049/191/2/352}

\bibitem[{{Koleva} {et~al.}(2011){Koleva}, {Prugniel}, {De Rijcke}, \&
  {Zeilinger}}]{Koleva2011}
{Koleva}, M., {Prugniel}, P., {De Rijcke}, S., \& {Zeilinger}, W.~W. 2011,
  \mnras, 417, 1643, \dodoi{10.1111/j.1365-2966.2011.19057.x}

\bibitem[{{Kormendy}(1985)}]{Kormendy1985}
{Kormendy}, J. 1985, \apj, 295, 73, \dodoi{10.1086/163350}

\bibitem[{{Kormendy} \& {Bender}(2012)}]{Kormendy2012}
{Kormendy}, J., \& {Bender}, R. 2012, \apjs, 198, 2,
  \dodoi{10.1088/0067-0049/198/1/2}

\bibitem[{{Kormendy} {et~al.}(2009){Kormendy}, {Fisher}, {Cornell}, \&
  {Bender}}]{Kormendy2009}
{Kormendy}, J., {Fisher}, D.~B., {Cornell}, M.~E., \& {Bender}, R. 2009, \apjs,
  182, 216, \dodoi{10.1088/0067-0049/182/1/216}

\bibitem[{{Kroupa}(2001)}]{Kroupa2001}
{Kroupa}, P. 2001, \mnras, 322, 231, \dodoi{10.1046/j.1365-8711.2001.04022.x}

\bibitem[{{Lee} {et~al.}(2006){Lee}, {Skillman}, {Cannon}, {Jackson}, {Gehrz},
  {Polomski}, \& {Woodward}}]{Lee2006}
{Lee}, H., {Skillman}, E.~D., {Cannon}, J.~M., {et~al.} 2006, \apj, 647, 970,
  \dodoi{10.1086/505573}

\bibitem[{{Leethochawalit} {et~al.}(2019){Leethochawalit}, {Kirby}, {Ellis},
  {Moran}, \& {Treu}}]{Leethochawalit2019}
{Leethochawalit}, N., {Kirby}, E.~N., {Ellis}, R.~S., {Moran}, S.~M., \&
  {Treu}, T. 2019, \apj, 885, 100, \dodoi{10.3847/1538-4357/ab4809}

\bibitem[{{Leethochawalit} {et~al.}(2018){Leethochawalit}, {Kirby}, {Moran},
  {Ellis}, \& {Treu}}]{Leethochawalit2018}
{Leethochawalit}, N., {Kirby}, E.~N., {Moran}, S.~M., {Ellis}, R.~S., \&
  {Treu}, T. 2018, \apj, 856, 15, \dodoi{10.3847/1538-4357/aab26a}

\bibitem[{{Lequeux} {et~al.}(1979){Lequeux}, {Peimbert}, {Rayo}, {Serrano}, \&
  {Torres-Peimbert}}]{Lequeux1979}
{Lequeux}, J., {Peimbert}, M., {Rayo}, J.~F., {Serrano}, A., \&
  {Torres-Peimbert}, S. 1979, \aap, 500, 145

\bibitem[{{Magrini} {et~al.}(2012){Magrini}, {Hunt}, {Galli}, {Schneider},
  {Bianchi}, {Maiolino}, {Romano}, {Tosi}, \& {Valiante}}]{Magrini2012}
{Magrini}, L., {Hunt}, L., {Galli}, D., {et~al.} 2012, \mnras, 427, 1075,
  \dodoi{10.1111/j.1365-2966.2012.22055.x}

\bibitem[{{Marigo} \& {Girardi}(2007)}]{Marigo2007}
{Marigo}, P., \& {Girardi}, L. 2007, \aap, 469, 239,
  \dodoi{10.1051/0004-6361:20066772}

\bibitem[{{Markwardt}(2012)}]{Markwardt2012}
{Markwardt}, C. 2012, {MPFIT: Robust non-linear least squares curve fitting}.
\newblock \doeprint{1208.019}

\bibitem[{{Martin} {et~al.}(2013){Martin}, {Ibata}, {McConnachie}, {Mackey},
  {Ferguson}, {Irwin}, {Lewis}, \& {Fardal}}]{Martin2013}
{Martin}, N.~F., {Ibata}, R.~A., {McConnachie}, A.~W., {et~al.} 2013, \apj,
  776, 80, \dodoi{10.1088/0004-637X/776/2/80}

\bibitem[{{Mayer} {et~al.}(2006){Mayer}, {Mastropietro}, {Wadsley}, {Stadel},
  \& {Moore}}]{Mayer2006}
{Mayer}, L., {Mastropietro}, C., {Wadsley}, J., {Stadel}, J., \& {Moore}, B.
  2006, \mnras, 369, 1021, \dodoi{10.1111/j.1365-2966.2006.10403.x}

\bibitem[{{McClure} \& {van den Bergh}(1968)}]{McClure1968}
{McClure}, R.~D., \& {van den Bergh}, S. 1968, \aj, 73, 313,
  \dodoi{10.1086/110634}

\bibitem[{{Mentz} {et~al.}(2016){Mentz}, {La Barbera}, {Peletier},
  {Falc{\'o}n-Barroso}, {Lisker}, {van de Ven}, {Loubser}, {Hilker},
  {S{\'a}nchez-Janssen}, {Napolitano}, {Cantiello}, {Capaccioli}, {Norris},
  {Paolillo}, {Smith}, {Beasley}, {Lyubenova}, {Munoz}, \& {Puzia}}]{Mentz2016}
{Mentz}, J.~J., {La Barbera}, F., {Peletier}, R.~F., {et~al.} 2016, \mnras,
  463, 2819, \dodoi{10.1093/mnras/stw2129}

\bibitem[{{Murray} {et~al.}(2005){Murray}, {Quataert}, \&
  {Thompson}}]{Murray2005}
{Murray}, N., {Quataert}, E., \& {Thompson}, T.~A. 2005, \apj, 618, 569,
  \dodoi{10.1086/426067}

\bibitem[{{Nagao} {et~al.}(2006){Nagao}, {Maiolino}, \& {Marconi}}]{Nagao2006}
{Nagao}, T., {Maiolino}, R., \& {Marconi}, A. 2006, \aap, 459, 85,
  \dodoi{10.1051/0004-6361:20065216}

\bibitem[{{Newman} {et~al.}(2013){Newman}, {Cooper}, {Davis}, {Faber}, {Coil},
  {Guhathakurta}, {Koo}, {Phillips}, {Conroy}, {Dutton}, {Finkbeiner}, {Gerke},
  {Rosario}, {Weiner}, {Willmer}, {Yan}, {Harker}, {Kassin}, {Konidaris},
  {Lai}, {Madgwick}, {Noeske}, {Wirth}, {Connolly}, {Kaiser}, {Kirby},
  {Lemaux}, {Lin}, {Lotz}, {Luppino}, {Marinoni}, {Matthews}, {Metevier}, \&
  {Schiavon}}]{Newman2013}
{Newman}, J.~A., {Cooper}, M.~C., {Davis}, M., {et~al.} 2013, \apjs, 208, 5,
  \dodoi{10.1088/0067-0049/208/1/5}

\bibitem[{{O'Sullivan} \& {Chen}(2020)}]{OSullivan2020b}
{O'Sullivan}, D., \& {Chen}, Y. 2020, arXiv e-prints, arXiv:2011.05444.
\newblock \doarXiv{2011.05444}

\bibitem[{{O'Sullivan} {et~al.}(2020){O'Sullivan}, {Martin}, {Matuszewski},
  {Hoadley}, {Hamden}, {Neill}, {Lin}, \& {Parihar}}]{OSullivan2020}
{O'Sullivan}, D.~B., {Martin}, C., {Matuszewski}, M., {et~al.} 2020, \apj, 894,
  3, \dodoi{10.3847/1538-4357/ab838c}

\bibitem[{{Panter} {et~al.}(2008){Panter}, {Jimenez}, {Heavens}, \&
  {Charlot}}]{Panter2008}
{Panter}, B., {Jimenez}, R., {Heavens}, A.~F., \& {Charlot}, S. 2008, \mnras,
  391, 1117, \dodoi{10.1111/j.1365-2966.2008.13981.x}

\bibitem[{{Pietrinferni} {et~al.}(2004){Pietrinferni}, {Cassisi}, {Salaris}, \&
  {Castelli}}]{Pietrinferni2004}
{Pietrinferni}, A., {Cassisi}, S., {Salaris}, M., \& {Castelli}, F. 2004, \apj,
  612, 168, \dodoi{10.1086/422498}

\bibitem[{{Revaz} \& {Jablonka}(2018)}]{Revaz2018}
{Revaz}, Y., \& {Jablonka}, P. 2018, \aap, 616, A96,
  \dodoi{10.1051/0004-6361/201832669}

\bibitem[{{Ruiz-Lara} {et~al.}(2018){Ruiz-Lara}, {Gallart}, {Beasley},
  {Monelli}, {Bernard}, {Battaglia}, {S{\'a}nchez-Bl{\'a}zquez}, {Florido},
  {P{\'e}rez}, \& {Mart{\'\i}n-Navarro}}]{Ruiz-Lara2018}
{Ruiz-Lara}, T., {Gallart}, C., {Beasley}, M., {et~al.} 2018, \aap, 617, A18,
  \dodoi{10.1051/0004-6361/201732398}

\bibitem[{{S{\'a}nchez-Bl{\'a}zquez} {et~al.}(2006){S{\'a}nchez-Bl{\'a}zquez},
  {Peletier}, {Jim{\'e}nez-Vicente}, {Cardiel}, {Cenarro},
  {Falc{\'o}n-Barroso}, {Gorgas}, {Selam}, \&
  {Vazdekis}}]{Sanchez-Blazquez2006}
{S{\'a}nchez-Bl{\'a}zquez}, P., {Peletier}, R.~F., {Jim{\'e}nez-Vicente}, J.,
  {et~al.} 2006, \mnras, 371, 703, \dodoi{10.1111/j.1365-2966.2006.10699.x}

\bibitem[{{Siegel} {et~al.}(2007){Siegel}, {Dotter}, {Majewski}, {Sarajedini},
  {Chaboyer}, {Nidever}, {Anderson}, {Mar{\'\i}n-Franch}, {Rosenberg}, {Bedin},
  {Aparicio}, {King}, {Piotto}, \& {Reid}}]{Siegel2007}
{Siegel}, M.~H., {Dotter}, A., {Majewski}, S.~R., {et~al.} 2007, \apjl, 667,
  L57, \dodoi{10.1086/522003}

\bibitem[{{Soderblom}(2010)}]{Soderblom2010}
{Soderblom}, D.~R. 2010, \araa, 48, 581,
  \dodoi{10.1146/annurev-astro-081309-130806}

\bibitem[{{Sybilska} {et~al.}(2017){Sybilska}, {Lisker}, {Kuntschner},
  {Vazdekis}, {van de Ven}, {Peletier}, {Falc{\'o}n-Barroso}, {Vijayaraghavan},
  \& {Janz}}]{Sybilska2017}
{Sybilska}, A., {Lisker}, T., {Kuntschner}, H., {et~al.} 2017, \mnras, 470,
  815, \dodoi{10.1093/mnras/stx1138}

\bibitem[{{Tremonti} {et~al.}(2004){Tremonti}, {Heckman}, {Kauffmann},
  {Brinchmann}, {Charlot}, {White}, {Seibert}, {Peng}, {Schlegel}, {Uomoto},
  {Fukugita}, \& {Brinkmann}}]{Tremonti2004}
{Tremonti}, C.~A., {Heckman}, T.~M., {Kauffmann}, G., {et~al.} 2004, \apj, 613,
  898, \dodoi{10.1086/423264}

\bibitem[{{Vargas} {et~al.}(2014){Vargas}, {Geha}, \& {Tollerud}}]{Vargas2014}
{Vargas}, L.~C., {Geha}, M.~C., \& {Tollerud}, E.~J. 2014, \apj, 790, 73,
  \dodoi{10.1088/0004-637X/790/1/73}

\bibitem[{{Vazdekis} {et~al.}(2016){Vazdekis}, {Koleva}, {Ricciardelli},
  {R{\"o}ck}, \& {Falc{\'o}n-Barroso}}]{Vazdekis2016}
{Vazdekis}, A., {Koleva}, M., {Ricciardelli}, E., {R{\"o}ck}, B., \&
  {Falc{\'o}n-Barroso}, J. 2016, \mnras, 463, 3409,
  \dodoi{10.1093/mnras/stw2231}

\bibitem[{Virtanen {et~al.}(2020)Virtanen, Gommers, Oliphant, Haberland, Reddy,
  Cournapeau, Burovski, Peterson, Weckesser, Bright, {van der Walt}, Brett,
  Wilson, Millman, Mayorov, Nelson, Jones, Kern, Larson, Carey, Polat, Feng,
  Moore, {VanderPlas}, Laxalde, Perktold, Cimrman, Henriksen, Quintero, Harris,
  Archibald, Ribeiro, Pedregosa, {van Mulbregt}, \& {SciPy 1.0
  Contributors}}]{2020SciPy-NMeth}
Virtanen, P., Gommers, R., Oliphant, T.~E., {et~al.} 2020, Nature Methods, 17,
  261, \dodoi{10.1038/s41592-019-0686-2}

\bibitem[{{Weisz} {et~al.}(2014){Weisz}, {Dolphin}, {Skillman}, {Holtzman},
  {Gilbert}, {Dalcanton}, \& {Williams}}]{Weisz2014}
{Weisz}, D.~R., {Dolphin}, A.~E., {Skillman}, E.~D., {et~al.} 2014, \apj, 789,
  147, \dodoi{10.1088/0004-637X/789/2/147}

\bibitem[{{Yates} {et~al.}(2020){Yates}, {Schady}, {Chen}, {Schweyer}, \&
  {Wiseman}}]{Yates2020}
{Yates}, R.~M., {Schady}, P., {Chen}, T.~W., {Schweyer}, T., \& {Wiseman}, P.
  2020, \aap, 634, A107, \dodoi{10.1051/0004-6361/201936506}

\bibitem[{{Young} \& {Lo}(1997)}]{Young1997}
{Young}, L.~M., \& {Lo}, K.~Y. 1997, \apj, 476, 127, \dodoi{10.1086/303618}

\bibitem[{{Zahid} {et~al.}(2013){Zahid}, {Geller}, {Kewley}, {Hwang},
  {Fabricant}, \& {Kurtz}}]{Zahid2013}
{Zahid}, H.~J., {Geller}, M.~J., {Kewley}, L.~J., {et~al.} 2013, \apjl, 771,
  L19, \dodoi{10.1088/2041-8205/771/2/L19}

\end{thebibliography}
\bibliographystyle{aasjournal}

\end{document}